\documentclass[11pt,a4paper]{article}
\usepackage{jcappub}

\title{Dark Matter Direct Search Rates in Simulations of the Milky Way and Sagittarius Stream}

\author{Chris W. Purcell,}
\author{Andrew R. Zentner,}
\author{and Mei-Yu Wang}

\affiliation{Department of Physics and Astronomy \& Pittsburgh Particle physics, Astrophysics and Cosmology Center (PITT PACC), University of Pittsburgh 15260 USA}

\emailAdd{cpurcell@pitt.edu}

\abstract{
We analyze self-consistent $N$-body simulations of the 
Milky Way disk and the ongoing disruption of the Sagittarius dwarf satellite 
to study the effect of Sagittarius tidal debris on dark matter detection experiments.  
In agreement with significant previous work, we reiterate that the standard halo model is 
insufficient to describe the non-Maxwellian velocity distribution of 
the Milky Way halo in our equilibrium halo-only and halo/galaxy models, 
and offer suggestions for correcting for this discrepancy.  
More importantly, we emphasize that the dark matter component of the 
leading tidal arm of the Sagittarius dwarf is significantly more 
extended than the stellar component of the arm, since the dark 
matter and stellar streams are not necessarily coaxial and may be 
offset by several kpc at the point at which they impact the Galactic disk.  
This suggests that the dark matter component of the Sagittarius debris 
is likely to have a non-negligible influence on dark matter detection 
experiments even when the {\em stellar} debris is centered several 
kpc from the solar neighborhood.  Relative to models without an infalling 
Sagittarius dwarf, the Sagittarius dark matter debris in our models 
induces an energy-dependent enhancement of direct search event 
rates of as much as $\sim 20-45\%$, an energy-dependent reduction 
in the amplitude of the annual modulation of the event rate by as 
much as a factor of two, a shift in the phase of the annual modulation 
by as much as $\sim 20$ days, and a shift in the recoil energy at 
which the modulation reverses phase.  These influences of Sagittarius 
are of general interest in the interpretation of dark matter searches, 
but may be particularly important in the case of relatively light 
($m_{\chi} \lesssim 20$~GeV) dark matter because the Sagittarius stream 
impacts the solar system at high speed compared to the primary halo dark matter.
}


\arxivnumber{1203.6617}

\begin{document}
\maketitle
\flushbottom

\section{Introduction} 

Numerous experiments aim to detect particle dark matter directly by 
measuring the rate of extremely rare nuclear recoil 
events from the elastic scattering of weakly-interacting massive particles 
(WIMPs; for context on this process, see \cite{goodman_witten85,jungman_etal96,bertone_etal05}). 
To interpret properly any putative direct detection signal or the limits on particle properties 
implied by any null searches, a detailed understanding of the phase space 
distribution of dark matter in the solar neighborhood is required (\cite{drukier_etal86,mccabe2010,lisanti_etal11}; 
see also \cite{green2011} for general review of the topic).  
Predictions for the rate of scattering events in direct search 
experiments are usually made under standard assumptions of a 
canonical local WIMP density, $\rho_0 \sim 0.3-0.4~\mathrm{GeV/cm^3}$, and a 
WIMP velocity distribution described by a Maxwellian function with a 
three-dimensional dispersion, $\sigma_{\mathrm{3D}} \sim 270 \pm 70$~km/s \cite{kamionkowski_98}.

Current experiments probe the parameter space of WIMP mass and scattering cross-section 
addressing many interesting dark matter candidates, including the lightest superpartner particles in 
supersymmetric theories \cite{bertone_etal12}.  Tentative signals suggest that the WIMPs comprising 
the cosmological dark matter may be particles with mass on the order 
of $m_{\chi} \sim 5-20~\mathrm{GeV/}c^2$ (e.g., the CRESST, DAMA/LIBRA, and CoGeNT 
experiments in \cite{angloher_etal11,bernabei_etal11,aalseth_etal11}, respectively).  
In this mass range, nuclear recoils with sufficient energy to be detectable require larger 
relative velocities than dark matter particle candidates in the 
higher mass range of $m_{\chi} \sim 100-1000~\mathrm{GeV/}$ that has 
been more widely explored in recent years 
(as demonstrated by Figure 1 in ref.~\cite{fox_etal11}).  The requirement of 
larger relative velocities renders scattering rates significantly more sensitive 
to the high-velocity tail of the dark matter velocity distribution than 
has been considered by many recent studies \cite{chang_etal09,hooper_kelso11}.  

The slight velocity anisotropy and significant spatiotemporal variation of 
the velocity distribution \cite{diemand_etal04,hansen_etal06,kuhlen_etal10} in a Milky Way-sized dark 
matter halo has been shown to affect significantly the predicted direct detection event rate 
compared to that obtained from a standard Maxwellian velocity distribution \cite{vergados_etal08,fairbairn_etal09}.  
The possibility of dark matter streams in the solar neighborhood has induced theoretical 
investigations of the effect of a coherent population of WIMPs on the annual modulation of the 
event rate observed by direct detection experiments 
\cite{freese_etal04,freese_etal05,savage_etal06,savage_etal07,savage_etal09,natarajan_etal11}.  
Meanwhile, high-resolution cosmological numerical simulations of halo formation 
have revealed that there are many kinematically-cold streams comprising the 
halo as well as myriad debris inflows from stripped subhalos 
\cite{kuhlen_etal12,lisanti_spergel11}.  These streams are associated with past mergers 
that built the Galactic mass to date, and are almost all of insufficient mass density 
to affect event rates measurably at the Earth's location in the Milky Way halo \cite{helmi_etal02,vogelsberger_etal09}.

Previous numerical studies that have addressed the issue of direct detection and 
velocity distributions in high-resolution $N$-body simulations 
(e.g., refs.~\cite{vergados_etal08,kuhlen_etal10}) have largely been limited 
to analyses of simulations including only dark matter.  These simulations 
do not account for the Milky Way disk and its cosmological growth and evolution 
to the present day.  Such experiments represent only individual samples of the 
broad statistical ensemble of merger histories that lead to the formation of a 
Milky Way-sized dark matter halo.  Consequently, these simulated halos lack 
specific structures known to exist in the Milky Way, such as the ongoing 
Sagittarius accretion and tidal stream, or other features that may have yet to be discovered.  
Therefore, the problem of mapping the results of 
cosmological numerical simulations onto specific predictions for the actual 
Milky Way halo, especially at the solar neighborhood, is a distinct challenge.  

Likewise, cosmological simulations designed to model the growth of 
a disk galaxy are not yet capable of accounting for the fundamental structures 
in a Milky Way-like galaxy.  Such simulations generally do not produce thin and 
dynamically-cold stellar disks analogous to that of the Milky Way (as demonstrated by 
ref. \cite{scannapieco_etal12} for a variety of computational techniques and energy feedback 
algorithms). Moreover, the computational expense of these campaigns imposes limits on the 
degree to which they can account for constraints on the unique mass assembly history of the Milky Way 
itself \cite{hammer_etal07,scannapieco_etal09}.  As an example, several hydrodynamical simulations have 
produced a Galaxy-sized host dark matter halo in a cosmological box, 
in concert with a significant, coherently-rotating ``dark disk'' component 
due to large accretions at late times \cite{read_etal09,ling_etal10,ling2010} -- 
although these components are widely found in a 
$\Lambda$CDM universe, the thin and cold Milky Way has had a quiescent 
accretion history that is incompatible with such recent minor mergers \cite{purcell_etal09b}.

Modeling the dark matter and luminous components of the Milky Way in equilibrium with 
each other, in an isolated simulation calibrated to specific characteristics of the 
Milky Way (as in the formalism of \cite{widrow_etal08}), complements cosmological simulations in a number of ways.  
Isolated simulations are presently rather well-constrained by observational 
data on Galactic structure and substructure and therefore may represent many aspects of the 
Milky Way more faithfully than an individual sample from a highly-stochastic ensemble of merger 
and accretion histories of Milky Way-sized halos in dark matter simulations that neglect baryons.
Such isolated calculations represent one method for estimating the influences of significant baryonic 
structures, such as the Galactic disk, and controlling for the unique merger history of the 
Milky Way.  Isolated simulations also provide a framework for incorporating important 
{\em accidental} features such as the Sagittarius dwarf galaxy (Sgr) and its associated 
tidal stream.  The disadvantage of this approach is that it does not preserve the 
cosmological construction of the host halo, in which virialized streams associated 
with high-redshift merger activity would be remnants of galaxies with so few stars 
that their present-day spatial distribution cannot be determined; 
these mergers are not self-consistently treated in an approach explicitly designed 
to model the current structure of the Milky Way, in equilibrium with a smooth dark matter 
host halo.  In the absence of good constraints on the late-time accretion history of the Galaxy, as determined 
observationally by mapping surveys, treating the Milky Way halo as an isolated and 
equilibrated disk/bulge/halo model at redshift $z=0$ is a complementary method 
to high-resolution cosmological simulations. 

\begin{figure}
\includegraphics[width=3.0in]{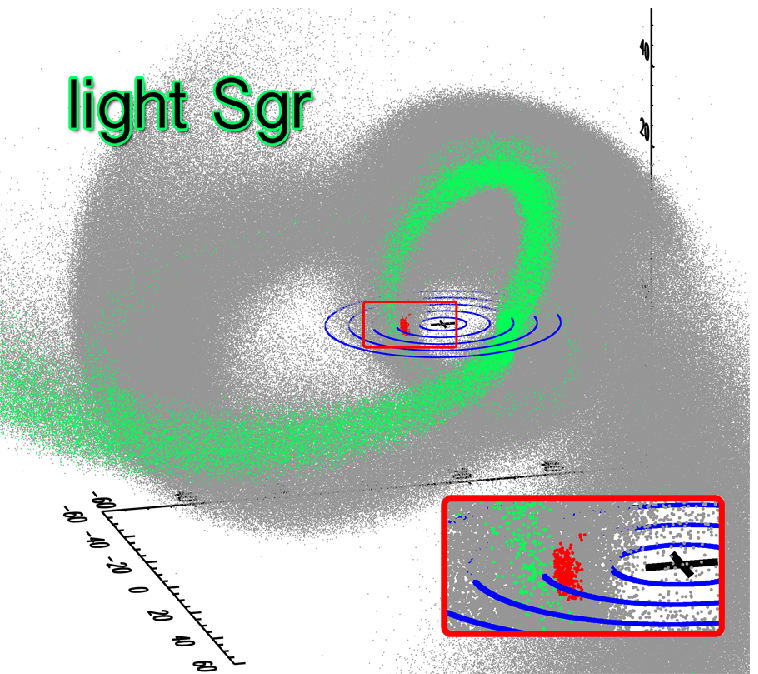}
\includegraphics[width=3.0in]{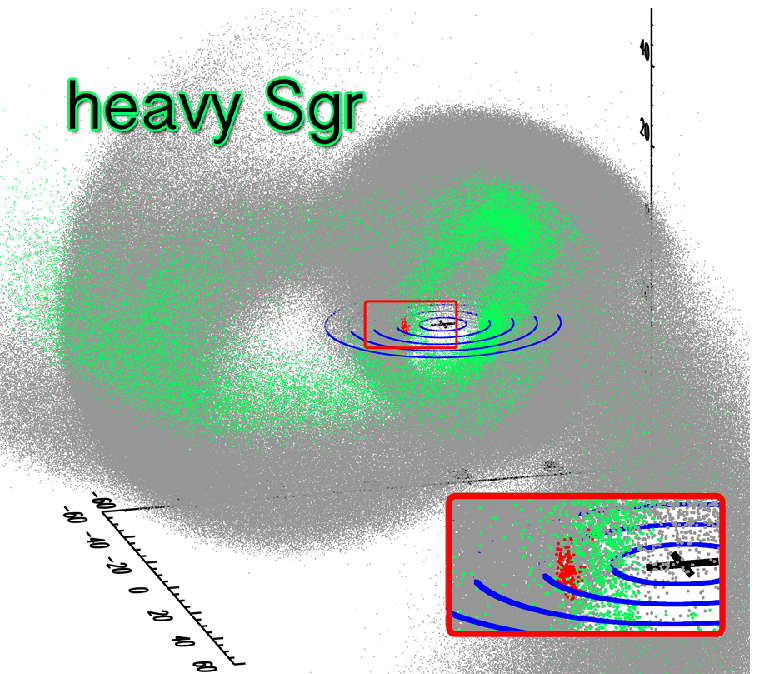}
\vspace{0.1in}
\includegraphics[width=6.09in]{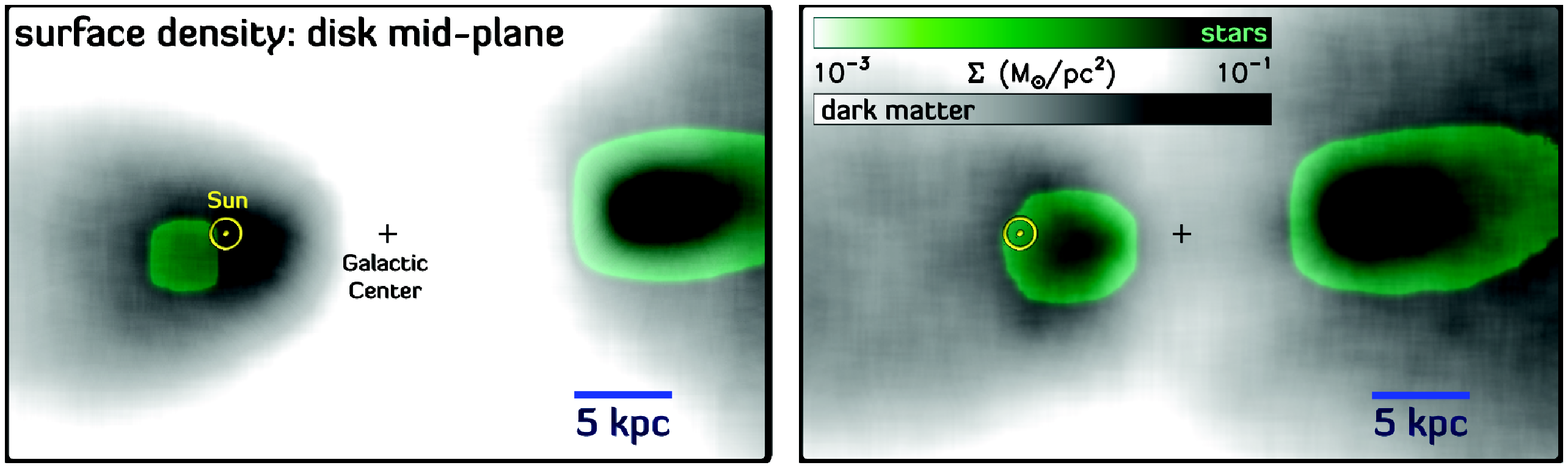}
\caption{{\em Upper} panels: distributions of stars and dark matter from the 
disrupting Sagittarius dwarf satellite galaxy ({\em green} and {\em gray} particles, respectively), 
in our {\em light Sgr} and {\em heavy Sgr} models.  In this perspective, the Milky Way disk 
plane is denoted by concentric blue rings at 5-kpc radial intervals, 
to a Galactocentric distance of 25 kpc.
{\em Lower} panels: surface density maps of Sgr stars and dark matter through the disk mid-plane 
for each model, computed in a slice with depth of 2 kpc, similar to the width of the Galactic disk.  
Note that the leading stellar stream does not pass directly through the solar 
neighborhood, although a significant amount of dark matter belonging to that stream 
is found near the Sun (as shown by {\em red} particles in the {\em upper} inset panels, and 
by grayscale shading in the {\em lower} panels).  Indeed, the projected contours of stellar 
density and dark matter density are not concentric, particularly in the more heavily-stripped 
{\em light Sgr} model.}
\label{fig:sgr}
\end{figure}

In this paper, we analyze a set of isolated simulations designed to represent the Milky Way 
disk along with the Sagittarius galaxy impact on the disk and the associated Sagittarius tidal 
debris.  We demonstrate that neglecting important modifications to $N$-body predictions, 
such as the presence of a large stellar Galactic disk and the ongoing Sagittarius dwarf merger, 
both structures that are known to exist, may well result in systematic misestimations of the expected 
event rates and annual modulation signatures in direct detection experiments.  
The stellar disk of the Milky Way itself, being in approximate equilibrium with its dark matter host halo, 
has concomitantly drawn the near-midplane region of the dark matter halo into a 
phase-space distribution peaked at higher velocity and with larger deviations from the Maxwellian 
form than found in dark matter-only simulations.  
These results follow from models of disk-halo equilibria formulated in ref.~\cite{widrow_etal08} and are 
related to numerous previous statements of uncertainty in the local dark matter density and velocity 
distribution in Milky Way models, as in the reconstructions of \cite{strigari_trotta09,salucci_etal10} among 
many other efforts.

Our most interesting results address the influence of the Sagittarius debris on dark matter detection 
experiments.  Some observational measurements of the tidally-stripped stars suggest that the 
leading {\em stellar} arm of the Sgr tidal stream may fall several kiloparsecs away from the 
solar position in the Galactic plane (e.g., \cite{belokurov_etal06,seabroke_etal08}); although recent 
detections of coherently-moving stellar populations in the solar neighborhood seem to rule out large-scale 
flows with Sagittarius-like vertical velocities \cite{helmi_etal99,re_fiorentin_etal11}, percent-level streams outside one 
or two kiloparsecs from the Sun are poorly constrained and difficult to disentangle using tracer-sampling techniques.  
Observationally-viable models for the Sgr infall exhibit several interesting features 
relevant to contemporary and future direct search experiments, because the models we present here 
treat the Sgr dwarf in a cosmological context, in the sense that we assume 
that the Sagittarius galaxy is itself embedded in a dark matter halo as it merges with the 
Milky Way.  Utilizing simulations of the Sagittarius infall, we emphasize that although the primary {\em stellar} component of the Sagittarius debris stream may be 
several kiloparsecs away, the associated {\em dark matter} stream is always significantly 
more spatially extended than the stellar stream (as shown in Figure~\ref{fig:sgr}) and 
more coherent in velocity space.  In fact, our models illustrate that the stellar and 
dark matter streams need not even be spatially concentric in realistic models.  In other words, 
the peak of the stellar density of the stream may be displaced from the peak in dark matter 
density associated with the stream by several kiloparsecs.  Both of these properties of 
simulated Sagittarius infall models suggest that it is not necessary for the solar neighborhood 
to be closer than several kiloparsecs away from the primary stellar stream of Sagittarius debris 
in order for the stream to affect significantly dark matter detection rates.  We use the dark matter 
streams realized in our simulations to model the dark matter velocity distribution in the solar 
neighborhood.  Indeed, we find that the near coincidence of the solar neighborhood and the debris 
of the Sagittarius satellite's dark matter halo creates a peak in the high-end of the dark matter 
velocity distribution.  These Sgr particles can lead to significant effects on the energy 
dependence of dark matter detection rates, as well as the amplitude and phase of the annual 
modulation of event rates, particularly for relatively low-mass dark matter candidates ($m_{\chi} \lesssim 20~\mathrm{GeV}$).

Our detailed calculations involve velocity distributions drawn from high-resolution numerical experiments, 
involving an isolated Galaxy-analog stellar disk/bulge/halo model and self-consistent treatment of the 
Sagittarius infall.  In \S\ref{sec:methods}, we discuss the relevant features of simulations 
introduced by \cite{purcell_etal11} and describe the analysis implying that significant 
alterations may be necessary to the predictions of event rates in direct detection experiments.  
Our results are presented in \S\ref{sec:results}, reserving \S\ref{sec:discuss} for discussion of 
their application and potential future prospects regarding detectors with low recoil-energy thresholds 
and more sensitive annual modulation constraints.

\section{Methods}
\label{sec:methods}

\subsection{Simulations of the Galactic Environment}

All analyses in this work are based on simulations described in \cite{purcell_etal11}, and 
we refer the reader there for technical details.  In brief, we 
describe four high-resolution experiments based on self-consistent 
halo/galaxy models, using throughout the notation ``{\em host halo}'' to refer to a standard NFW dark matter halo 
\cite{navarro_etal96} with a mass profile roughly consistent with that of the Milky Way, 
and ``{\em halo}$~+~${\em disk}'' for the model with an {\em identical} halo generated in equilibrium 
with a Galactic-analogue disk and central bulge with a global density structure consistent with 
that of the observed Milky Way stellar disk \cite{widrow_etal08}, and a total-mass surface density at the solar location that is 
constrained to match the observations of ref.~\cite{kuijken_gilmore91} and therefore consistent with recent refinements 
to this value \cite{holmberg_flynn04}.  
For the purposes of the present paper, in the context of phase-space behavior on 
small scales, we note that this three-component model represents an approximate, 
self-consistent solution to the coupled collisionless Boltzmann and Poisson equations, 
and that all {\em N}-body simulations were performed using many millions of particles 
with a force-softening resolution equal to one parsec.  The disk is stable against 
large-scale perturbations, and develops only a weak bar in isolated evolution over a 
timescale of $3-5$~Gyr.  The disk remains globally static in density and velocity 
structure outside a few kiloparsecs from the Galactic center, ensuring that the solar 
neighborhood in this model remains unaffected by artificial baryonic asymmetries related to spiral-arm evolution.  
  
Here and throughout, we define the solar neighborhood as a wedge in the radial 
range $8<R_{\odot}<9$~kpc and extending $\pm 2.5~(1.5)$~kpc in the vertical (tangential) direction.  
Wedges of this size are used to define the {\em kinematic} properties of the halo and stream dark 
matter particles because a large number of particles is needed to probe velocity structure.  
The relative densities of the dark matter and stream components can be determined within 
significantly smaller volumes.  In each case, we have verified that we derive consistent 
results when we vary the size of the wedge by $\sim 1-2~\mathrm{kpc}$ in the tangential 
direction and up to $\sim 10$~kpc in the vertical direction.  The insensitivity of our results follows from the 
relatively mild spatial variations in stream density exhibited in Figure~\ref{fig:sgr}.  
Following the example of ref.~\cite{chang_etal09}, the rest-frame on Earth is taken to include 
the galactic rotation (a purely tangential $v_c = 220$~km/s in the {\em host halo} model, and calculated 
according to the stellar disk's rotation in the {\em halo}$~+~${\em disk} and related models), 
as well as the peculiar solar motion $(U_{\odot},V_{\odot},W_{\odot}) = (11.1,12.2,7.3)$~km/s according to 
\cite{sbd_2010}, and finally the Earth's orbital motion around the Sun as prescribed by \cite{lewin_smith96}.  
The position of the impact of the Sagittarius stream on the Galactic disk is somewhat uncertain, 
so there is some arbitrariness in our identification of the solar neighborhood.  We identify the solar 
neighborhood such that the Sun has approximately the correct position relative to the Sagittarius 
dwarf remnant, and we return to uncertainty in the relative position of the Sun with respect to the stream in 
\S~\ref{subsec:sgr}.

As in \cite{purcell_etal11}, we examine two numerical experiments in which 
the Galactic disk is impacted by cosmologically-realistic progenitors to the 
Sagittarius dwarf galaxy, resulting in significant spirality and ring-like 
structure in the host Milky Way.  The dark matter halo masses of the two satellite 
models roughly bracket the expected range motivated by cosmological abundance 
matching techniques \cite[e.g.,][]{conroy_wechsler09}: ``{\em light Sgr}'' 
assumes a Sagittarius progenitor mass of $M_{\mathrm{Sgr}} = 10^{10.5} M_{\odot}$, 
while ``{\em heavy Sgr}'' has a progenitor mass of $M_{\mathrm{Sgr}} = 10^{11} M_{\odot}$.  
This cosmologically-plausible range is supported by the dynamical reconstruction of the 
progenitor as reported in ref.~\cite{niederste-ostholt_etal10}, in which the mass immediately prior to 
tidal disruption is estimated to be at least $\sim 10^{10} M_{\odot}$, a value consistent with the total 
masses contained within and truncated at the two model subhalos' initial Jacobi tidal radii when our 
simulations begin.  As reported in \cite{purcell_etal11}, both models are consistent with the approximate 
characteristics of the Sagittarius tidal debris as mapped by the Two-Micron All Sky Survey and the 
Sloan Digital Sky Survey among other efforts \cite{majewski_etal03,majewski_etal04,correnti_etal10}.  
In particular, these evolved debris distributions reproduce the spatial trends in observed radial velocity and 
heliocentric distance, although we note that our goal in this work is not to model the Sagittarius stellar stream 
in precise detail.  The massive dark matter component is disrupted on scales much larger than 
that of the systematic errors involved in both measuring and modeling the luminous debris, such that the 
general behavior of Sgr dark matter is grossly similar in both experiments, as we now address.

The presence of a coherent stream of dark matter in the solar neighborhood is a robust 
prediction of {\em both} Sagittarius models, and would be a feature of any putative progenitor with cosmologically-motivated and 
observationally-consistent properties as discussed above.  The material is stripped from the 
infalling satellite, and in both models the dark component of the leading tidal arm 
contributes a non-negligible fraction of dark matter to the solar neighborhood.  We emphasize that 
the contribution of dark matter from Sgr is non-negligible despite the fact that the 
{\em stellar arm} of the tidal debris is displaced from the solar neighborhood by several 
kiloparsecs in the Milky Way disk plane in our baseline models.  This occurs for two reasons.  
First, in any model in which the Sagittarius progenitor has a cosmologically-motivated 
progenitor halo, the progenitor halo is significantly larger than the stellar 
component of the progenitor galaxy and spawns a tidal stream that is significantly 
more spatially extended than the stellar debris stream.  Indeed, the dark matter 
streams in our models are more than $\gtrsim 10~\mathrm{kpc}$ in cross-sectional diameter.  
Second, the bulk of the dwarf's dark matter is stripped from the Sagittarius progenitor prior to the onset of stellar 
mass loss.  The dark matter and stellar debris material do not follow precisely identical orbits, so a 
cross-section of the dark matter stream at the plane of the 
disk is {\em not necessarily} concentric with a cross-section of the stellar debris.  
These features are both evident in Fig.~\ref{fig:sgr}, with the offset between the 
dark matter and stellar arms most prominent in the {\em light Sgr} model, since the progenitor 
in that case has been more heavily disrupted.

\begin{figure}
\includegraphics[width=3.02in]{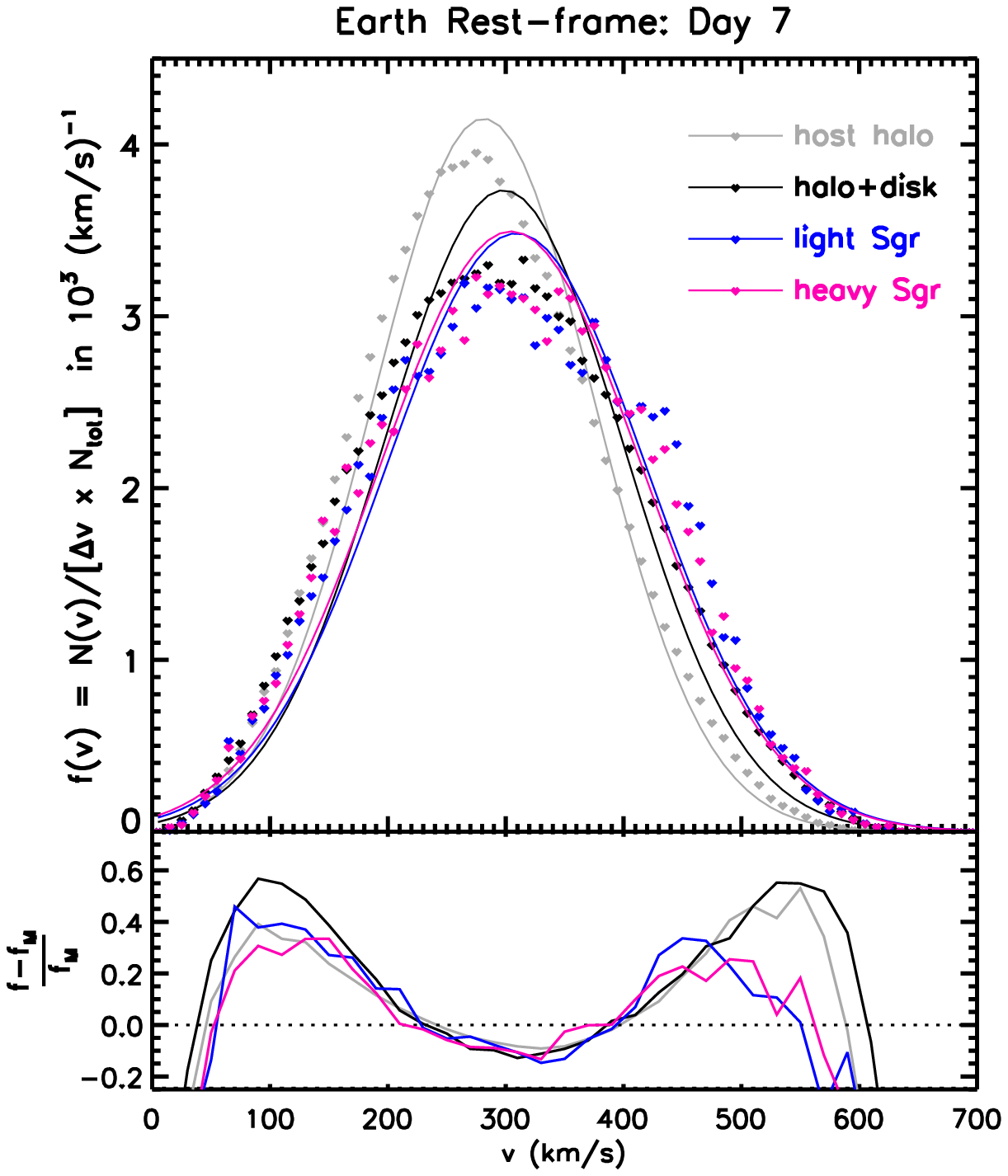}
\hspace{0.05in}
\includegraphics[width=2.87in]{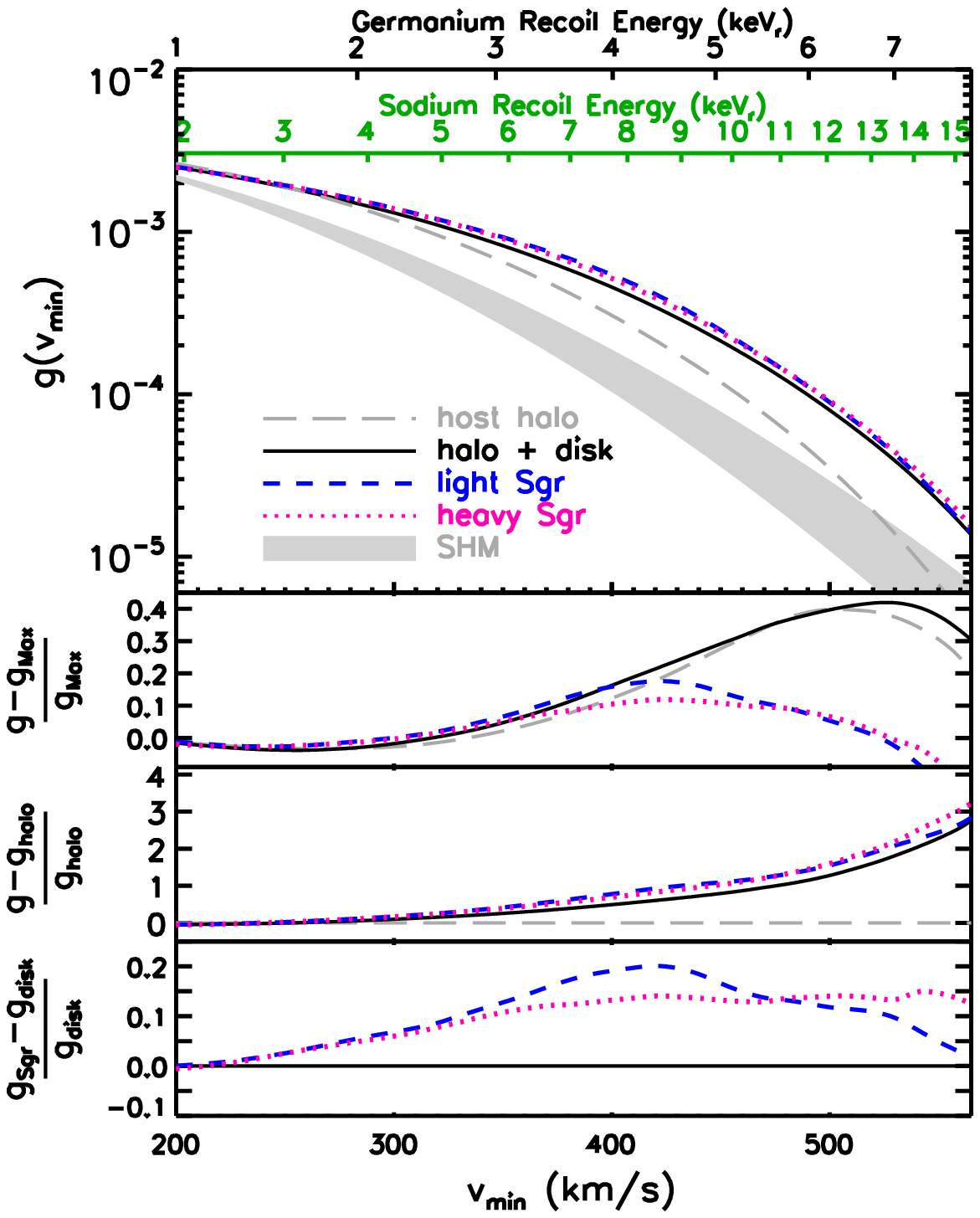}
\caption{{\em Left} --- Dark matter velocity distribution relative to Earth in the early 
Northern winter (day-number 7 relative to noon on December 31st, for the J2000.0 epoch), 
for each of our four models.  
The dotted histograms are best fit to a Maxwellian form by the 
{\em thin solid} lines in the {\em upper} panel, while the fractional 
differences from the Maxwellian fits are shown as a function 
of velocity in the {\em lower} panel. The excesses near $v \sim 450-500$~km/s 
in both of the Sagittarius models are real, and drift during the year from this 
wintertime peak to a summertime low of $v \sim 400$~km/s.  
{\em Right} --- The integral quantity $g(v_{\mathrm{min}})$ as a function of the 
minimum particle velocity, $v_{\mathrm{min}}$, required to initiate nuclear recoil in a direct detection 
experiment for each of our four models ({\em upper} panel; also shown is a comparison to 
the standard halo model (SHM) Maxwellian for a range of circular velocities $v_c=200-220$~km/s).  
The second panel from the top shows the residual of each model compared to a {\em best-fit} 
Maxwellian ({\em not} the SHM), the third panel from the top shows the residual of the 
models that include the Galactic disk compared to the {\em host halo} model, and the 
bottom panel shows the residual of the Sagittarius models compared to the {\em halo}$~+~${\em disk} 
model.  The integral $g(v_{\mathrm{min}}$) is directly proportional 
to the event rate so the generally higher velocities in the models including the 
Galactic disk result in a factor of roughly two 
or more increase in event rate relative to the {\em host halo} model. 
The features induced by each Sgr stream model lead to further modifications 
of this prediction on the $\sim 10-20\%$ level ({\em lower} panel).
For illustration, the alternate axes in the {\em upper} panel 
show corresponding nuclear recoil energies in 
unquenched keV$_{\mathrm{r}}$ for germanium and sodium target nuclei, 
assuming a WIMP mass of $m_{\chi} = 10$~GeV$/c^2$.
}
\label{fig:veldistg}
\end{figure}

Our two Sagittarius models shown in Fig.~\ref{fig:sgr} could in principle underestimate the amount of Sgr dark 
matter near the Sun, because {\em the axis of the leading stellar stream misses the solar neighborhood entirely in each case}, 
with only the outskirts of the dark matter in that tidal arm contributing to a potential WIMP detection signal.  
In fact, previous models for the Sagittarius leading arm have indicated that it may impact the disk plane 
within a few kpc of the Sun \cite{law_etal05}, although observational explorations of the 
solar neighborhood have failed to detect any sub-population kinematically similar to streaming substructure 
with Sagittarius-like geometry within one or two kiloparsecs of the Sun \cite{helmi_etal99,re_fiorentin_etal11}, and constraints on the descending portion of the 
leading arm are presently very poor (e.g., \cite{correnti_etal10}; see also the ``field of streams" depicted in ref.~\cite{belokurov_etal06}).  
We note that the more recent investigation in ref.~\cite{law_etal10} 
of the Sgr stream in a triaxial halo suffers from a number of flaws, including the lack of a dark matter component in the progenitor as 
well as the adoption of a static Galactic potential.  In contrast, our modeling techniques and initial conditions self-consistently resolve these 
issues in a numerical context that correctly treats the evolution of tidal disruption characteristics in the stellar debris.  As a further point of 
distinction, we also emphasize that the preferred model of ref.~\cite{law_etal10} (in which the leading arm passes $\sim 10$ kpc from the Sun) 
requires a nearly-oblate Milky Way halo (at odds with the prolate shape preferred by long-lived warps in the atomic-hydrogen 
gas layer in the Milky Way \cite{banerjee_jog11} as well as the Holmberg effect of satellites being clustered around a plane normal 
to that of the central galaxy \cite{helmi04}) with an unusual orientation compared to the cosmological findings of 
refs.~\cite{bailin_steinmetz05,debattista_etal08,bett_etal10}.  In any case, the orbital shape of the satellite's infall is sufficiently 
well-constrained that any debris model with a cosmologically-realistic dark subhalo will necessarily result in a wide stream of WIMPs 
raining into the solar system.


\subsection{Dark Matter Direct Detection Event Rates}

After obtaining the speed distributions characterized by $f(v)$ in the {\em left} panel of Figure~\ref{fig:veldistg}, 
for each of the four models we calculate differential event rates as a function of recoil energy as 
in \cite{jungman_etal96}:
\begin{equation}
\frac{dR}{dE_{\mathrm{r}}} (E_{\mathrm{r}},t) = \frac{{\sigma_{\chi}\rho_0}}{2\mu^2m_{\chi}} A^2 F^2(E_{\mathrm{r}}) \times g(v_{\mathrm{min}}),
\label{eqn1}
\end{equation}
where $\sigma_{\chi}$ is the WIMP cross-section for scattering on a proton (assuming here that the WIMP couples nearly equally to protons and neutrons), 
$\rho_0$ is the WIMP density in the solar neighborhood, $\mu$ is the reduced mass of the proton and a WIMP with 
mass $m_{\chi}$, $A$ is the atomic mass number of the detector nuclei, $F(E_{\mathrm{r}})$ is 
the form factor of nuclear scattering as a function of recoil energy $E_{\mathrm{r}}$, and the quantity $g(v_{\mathrm{min}})$ 
is the integral in velocity space of the velocity distribution divided by the WIMP speed, 

\begin{equation}
g(v_{\mathrm{min}}) = \int^\infty_{v_{\mathrm{min}}} \frac{f(v)}{v} dv .
\end{equation}
Here, $v_{\mathrm{min}} = \sqrt{\frac{E_{\mathrm{r}} M_a}{2 \mu_A^2}}$, where $M_a$~and $\mu_A$ are the atomic mass and WIMP-nucleon reduced mass, and this is the minimum relative speed 
necessary for nuclear recoil to yield an energy $E_{\mathrm{r}}$ (lower WIMP masses require higher values of $v_{\mathrm{min}}$ at fixed recoil energy).  In our analysis, we calculate 
$g(v_{\mathrm{min}})$ using Earth rest-frame velocity distributions and the local WIMP density $\rho_0$ directly from our simulations.  We adopt the Helm form factor $F(E_{\mathrm{r}})$ (as in 
\cite{kelso_hooper11}; see \cite{duda_etal07} for details and fitted parameters).  Throughout, we 
choose an arbitrary value of $\sigma_{\chi} = 10^{-40}~\mathrm{cm}^2$ ($=10^{-4}$~pb) for 
absolute event rates.  Predicted event rates can be scaled linearly for different values of 
$\sigma_{\chi}$.  We frame our results in terms of unquenched nuclear recoil energy $E_{\mathrm{r}}$, 
using the standard unit notation keV$_{\mathrm{r}}$, reminding the reader that 
the quenching factor for a particular detector material must be used 
to convert this to electron-equivalent recoil energy 
in keVee, i.e. $E_{\mathrm{keVee}} = qE_{\mathrm{keVr}}^x$.  
For the detector examples we investigate in this work, 
$(q,x) \simeq (0.199,1.12)$~for germanium \cite{mccabe2011}, and 
$(q,x) \simeq (0.3,1.0)$ for sodium \cite{bernabei_etal11}.  Note that we do not model the 
finite energy resolution of detectors.

 \section{Results}
\label{sec:results}

\subsection{General Deviations from the Standard Halo Model}

Our primary results concern the non-trivial amount of dark matter donated to the solar neighborhood by 
the leading tidal arm of the Sagittarius dwarf galaxy (shown in Figure~\ref{fig:sgr}), but prior to 
this discussion, we briefly itemize pertinent features of the dark matter component of the primary 
halo emphasizing deviations from a Maxwellian distribution.  Many of the gross features that 
characterize deviations from a Maxwellian have been pointed out in previous work 
\cite{diemand_etal04,hansen_etal06,kuhlen_etal10,vergados_etal08,fairbairn_etal09}.  
We note that the particular systems we study 
have been tuned to equilibrium {\em host halo} or {\em halo}$~+~${\em disk} configurations 
that represent many of the gross features of the Milky Way galaxy, but that these are 
not unique solutions for equilibrium models of the Milky Way.  
Dark matter direct search rates are usually 
interpreted in the context of a standard halo model (SHM) 
\cite{chang_etal09,hooper_kelso11} with $\rho(r) \propto r^{-2}$, 
a local density of $\rho_0 = 0.3~\mathrm{GeV}/\mathrm{cm}^{3}$, 
and a velocity distribution described by a Maxwellian form with a 
three-dimensional dispersion $\sigma_{\mathrm{3D}}=\sqrt{3/2}v_c$, 
where $v_c$ is the circular speed at the solar radius which is assumed to be identical to the 
peak speed of the Maxwellian distribution.  Fig.~\ref{fig:veldistg} shows that 
the equilibrium models we consider result in direct search rates 
significantly different from the specific Maxwellian assumed to derive 
SHM predictions (the shaded bands in the {\em upper right} panel).  Furthermore, the velocity 
distributions deviate significantly not only from the SHM, but 
from the general form of a Maxwellian speed distribution. 
Fig.~\ref{fig:veldistg} shows comparisons 
between velocity distributions in our equilibrium galaxy models and Maxwellian 
distributions with mean and dispersion that best fit the simulation data.  
The isolated {\em host halo} exhibits significant deviations from its best-fit 
Maxwellian distribution as shown in Fig.~\ref{fig:veldistg} and noted in 
previous studies and in agreement with the cosmological $N$-body results 
presented by ref.~\cite{kuhlen_etal10}.  The same is true for the {\em halo}$~+~${\em disk} and Sgr-infall models.  
In both cases best-fit Maxwellians underestimate the value of 
$g(v_{\mathrm{min}})$~by $\sim 20-40\%$ over a wide range of recoil energies relevant 
to direct detection experiments.  

\begin{figure}
\includegraphics[width=3.0in]{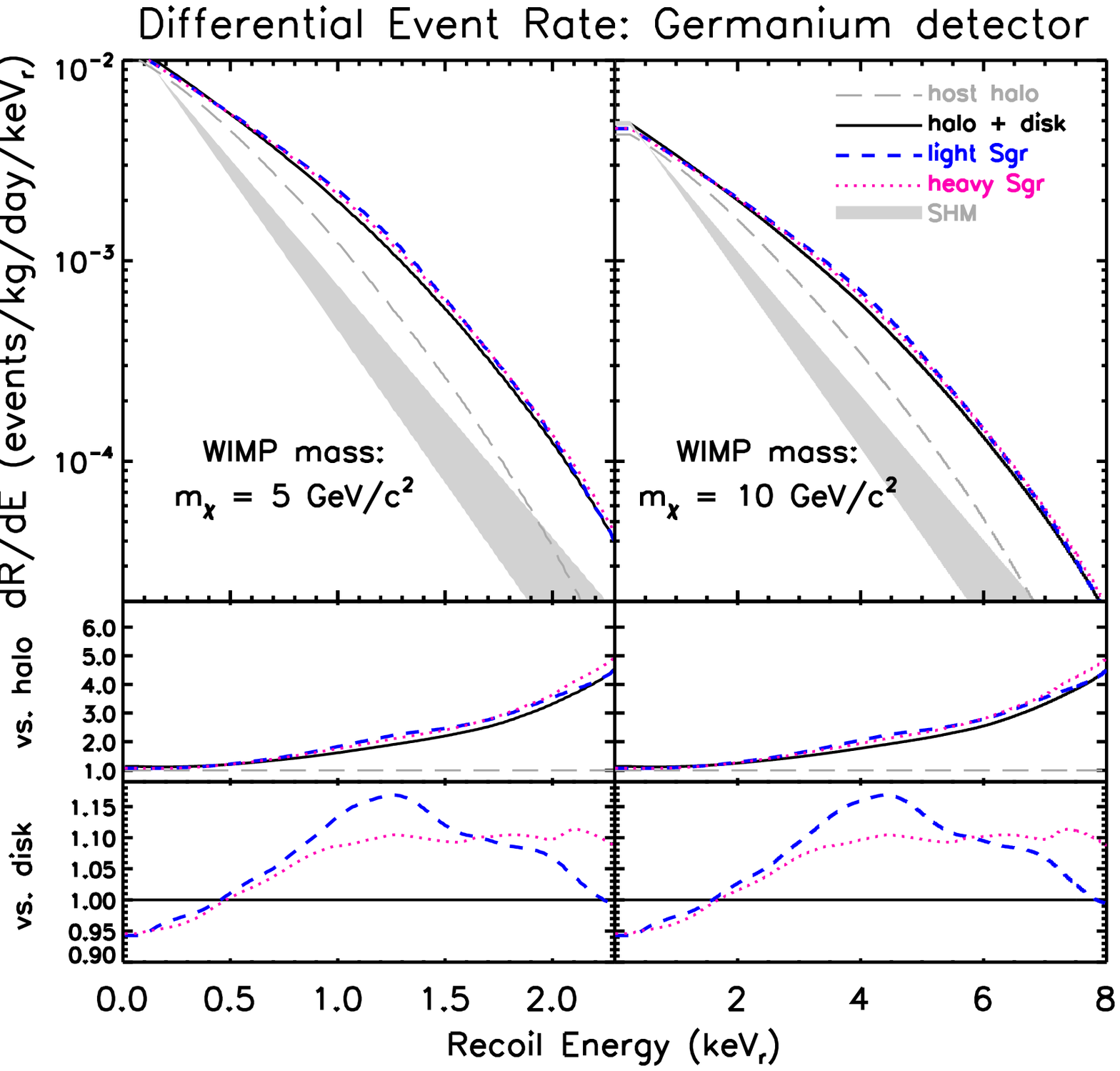}
\hspace{0.1in}
\includegraphics[width=2.98in]{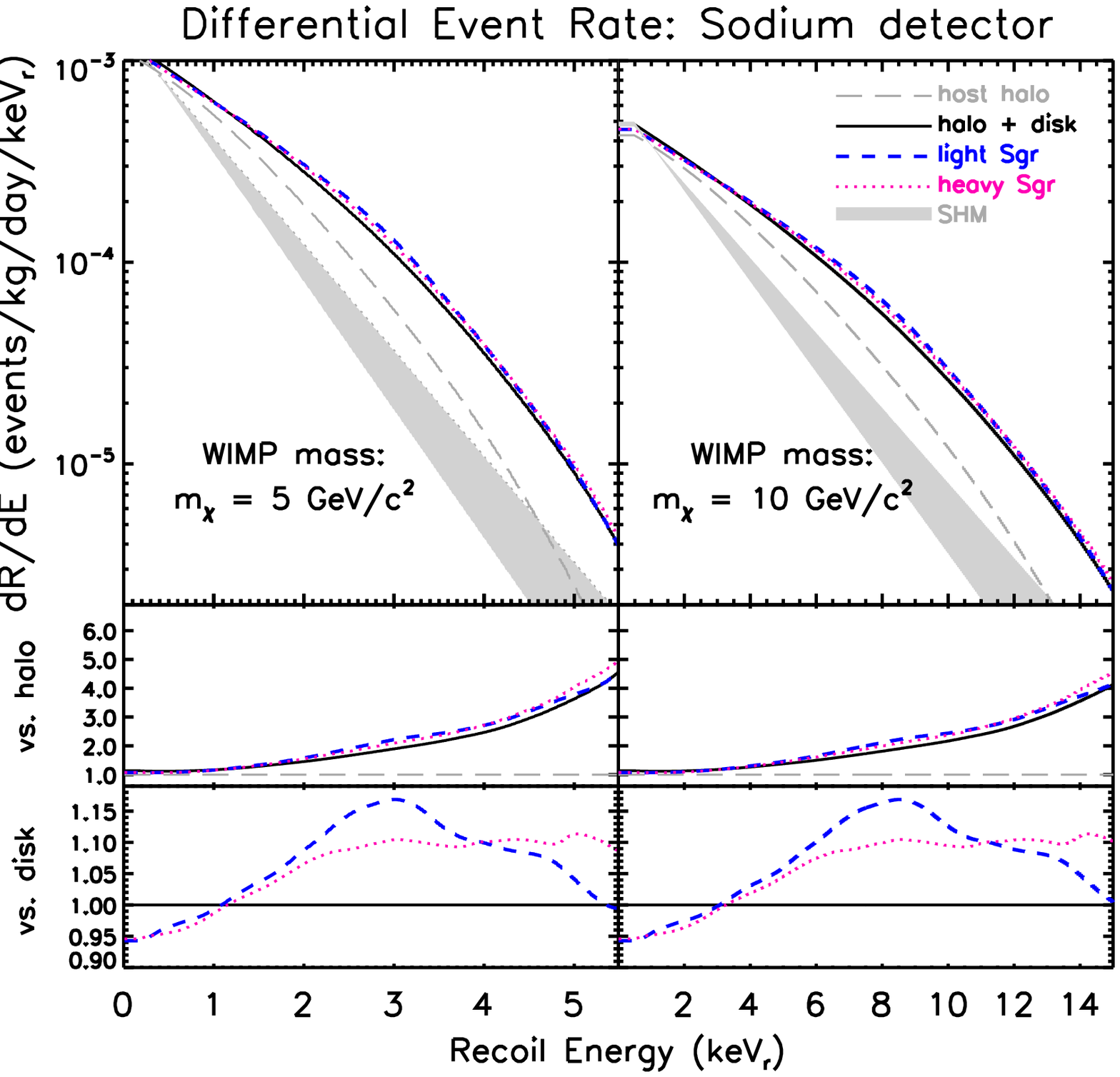}
\caption{Examples of predicted direct search event rates.  We show 
time-averaged differential scattering event rate $dR/dE$ as a function of 
recoil energy (keV$_{\mathrm{r}}$, unquenched), for scattering off of germanium 
({\em left} panels) and sodium ({\em right} panels).  
In each detector, we show example spectra for light WIMPs of mass $m_{\chi}=5$~and $10$~GeV/$c^2$.  
We show the fractional deviation in the event rate 
from the {\em host halo} model in the {\em center} panels, and the 
fractional deviation of the Sagittarius model event rates from 
the {\em halo}~+~{\em disk} values are shown in the {\em lower} panels.
}
\label{fig:drde}
\end{figure}

Relative to an isolated halo, the addition of the Galactic disk causes the host halo 
to contract globally as well vertically toward the disk mid-plane.  
The global contraction of the host halo in response to a disk shifts 
the velocity distribution to larger velocities (Fig.~\ref{fig:veldistg}), 
resulting in relatively higher event rates (Figure~\ref{fig:drde}).  
In addition to this gross shift, equilibrium Galaxy models including a 
disk exhibit speed distributions with significantly more broad, flat peaks.  
The kurtosis of the {\em host halo} speed distribution is $K\approx-0.3$, while the 
galaxy models have $K\approx-0.5$ (by definition $K=0$ for a Gaussian distribution, and $K \simeq 0.1$ for a Maxwellian).  
Scattering event rates are directly proportional to the local WIMP density $\rho_0$, 
which in all cases is higher for simulations including 
the Milky Way stellar disk than for the dark-matter-only {\em host halo} model.  
The solar neighborhood in this original halo resides in a region with WIMP density 
$\rho_0 \sim 0.53~\mathrm{GeV/cm^3}$, while the three models including the stellar 
disk of the Milky Way have $\rho_0 \sim 0.61-0.63~\mathrm{GeV/cm^3}$, a 
$\sim 20\%$ increase due to the overall contraction of the halo in response to the 
Galactic potential and the compression of the halo in the vertical direction due to 
the planarity of the disk.  We note that our models have been tuned to Milky Way structural properties 
\cite{widrow_etal08} (and also respond to variously-sized satellite impacts in a fashion consistent with 
that of the quiescent Galactic mass accretion history and the global spiral structure of the disk; see 
refs.~\cite{purcell_etal11,purcell_etal09}), so these values of $\rho_0$ illustrate the potential error 
latent in typical choices of this normalization.  
We present event rates calculated self-consistently with 
respect to the WIMP density $\rho_0$ in each model; however, 
the changes in the speed distribution and, therefore, the 
integrated quantity $g(v_{\mathrm{min}})$, represent far more 
significant alterations to predicted event rates.  Figure~\ref{fig:veldistg} 
and Figure~\ref{fig:drde} show that the shift in the speed distribution 
increases $g(v_{\mathrm{min}})$, and thus event rates, by a factor of several 
at high $v_{\mathrm{min}}$, a boost that is particularly important for light 
WIMP masses $m_{\chi} \sim 5-20$~GeV$/c^2$ suggested by several recent 
experiments \cite{angloher_etal11,bernabei_etal11,aalseth_etal11}.

Unsurprisingly, the gross effect of the Galactic component is to deepen the 
gravitational potential, resulting in generally larger relative speeds of 
dark matter particles.  It is interesting to assess whether this offset can 
be modeled simply so that they may be incorporated into future analyses, 
such as those proposed by ref.~\cite{peter2011} among others.  A simple proposal would be to employ 
a model for adiabatic halo contraction  \cite{blumenthal_etal86,gnedin_etal04} 
on the original {\em host halo}, which 
is well described by the standard NFW profile form.   
Specifying a velocity distribution would still be a challenge, 
but a simple proposal would be to employ the standard Eddington relation 
on the contracted halo (see the relevant exercises in ref.~\cite{binney_tremaine08}, for example).  
Such an approach would certainly not be 
self-consistent as both formalisms assume spherical symmetry and the 
Eddington relation does not yield a unique speed distribution; however, 
it is interesting to explore such an option as a simple, practical alternative 
to perform a gross correction to account for contraction without the expense 
of constructing equilibrium models of the Galaxy and its halo.\footnote{Among other caveats to such a formalism, some numerical experiments 
of Galaxy-scale halos have indicated that halos may expand as 
baryonic feedback ejects material from the galaxy \cite{governato_etal12,maccio_etal12}, 
working against the tendency toward adiabatic 
contraction in a way not easily modeled by analytic forms.}  
We find that our models can be described in this way at a similar level of 
precision as using the best-fit Maxwellian for each model.  In particular, 
we find residuals similar to the {\em halo}$~+~${\em disk} residuals for the best-fit 
Maxwellian in Figure~\ref{fig:veldistg}, with adiabatic contraction modeling 
resulting in residuals from $\sim 20-50\%$ for $v_{\mathrm{min}} \gtrsim 300$~km/s.  
At this level of precision, a simple analytical correction may be able to map results from 
a host halo identified in a dark-matter-only simulation to event rates in the 
solar neighborhood.  An even more parsimonious exercise recovers the halo/galaxy speed distribution 
by scaling each particle's velocity in the {\em host halo} by an amount equal to the increase 
in the distribution's peak speed; for our models, this simple adjustment produces event rates within $\sim 5\%$ of 
those yielded by the {\em halo}$~+~${\em disk} analysis.

\begin{figure}
\begin{center}
\includegraphics[width=2.6in]{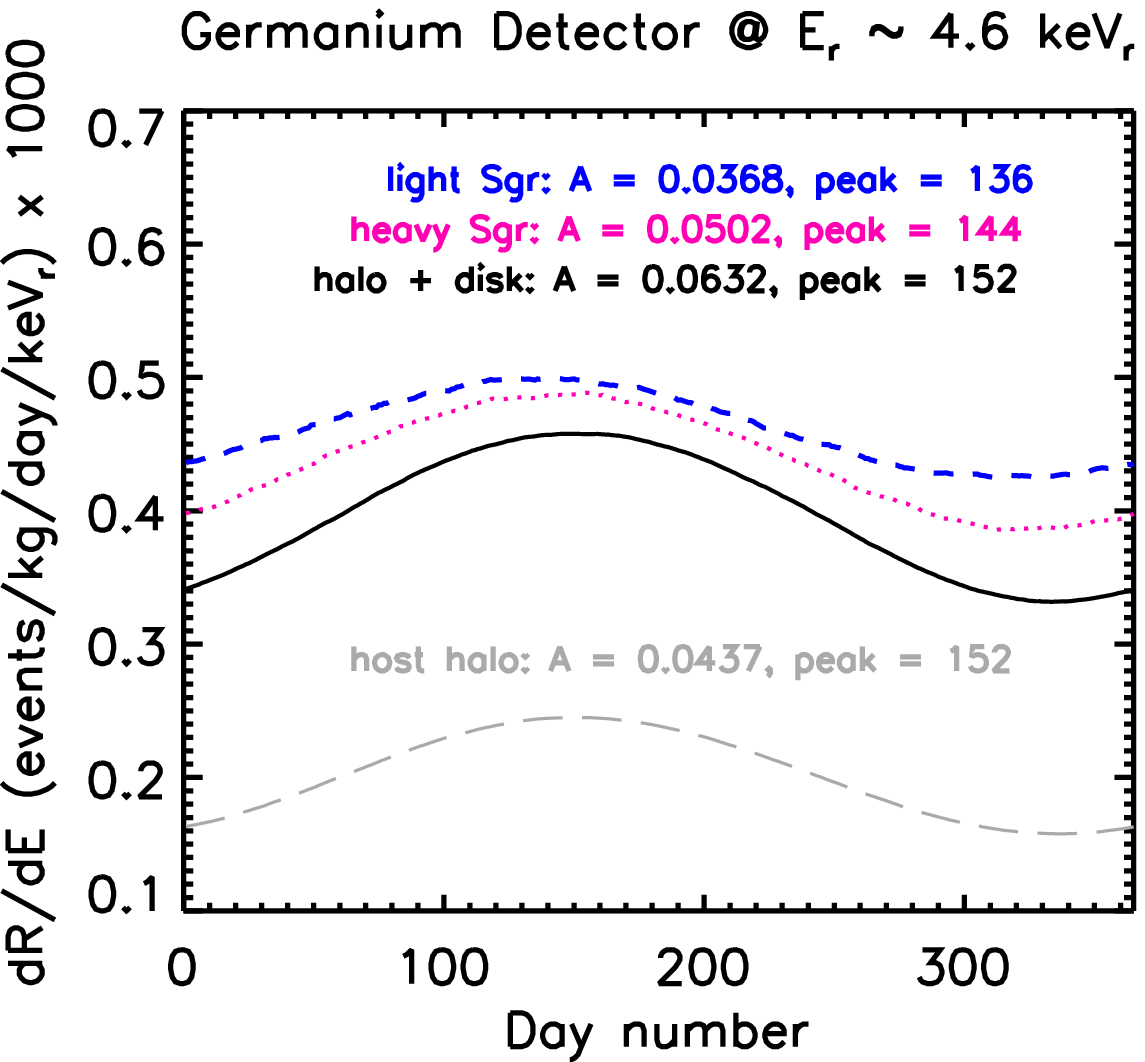}
\hspace{0.1in}
\includegraphics[width=2.7in]{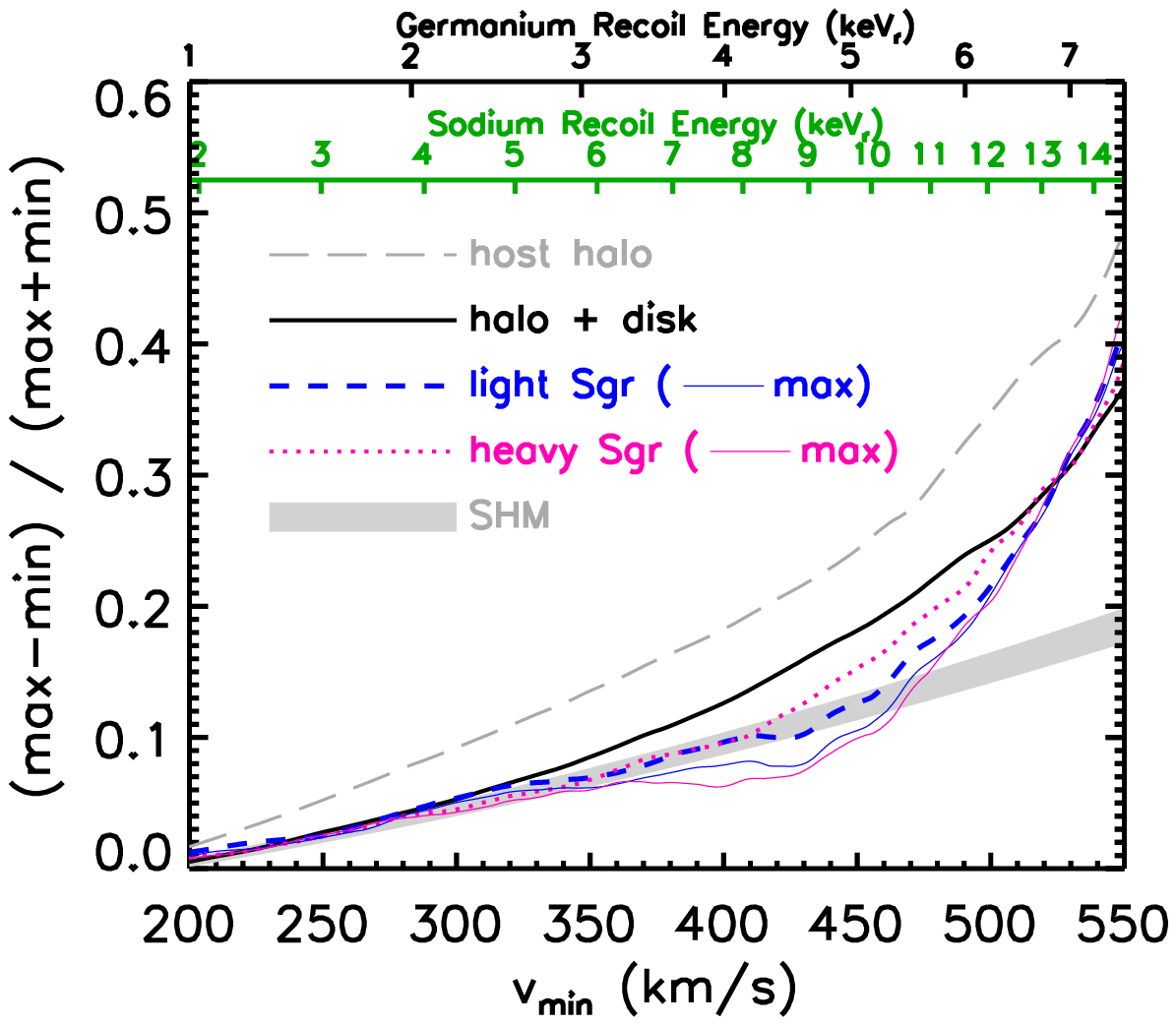}
\end{center}
\begin{center}
\includegraphics[width=2.71in]{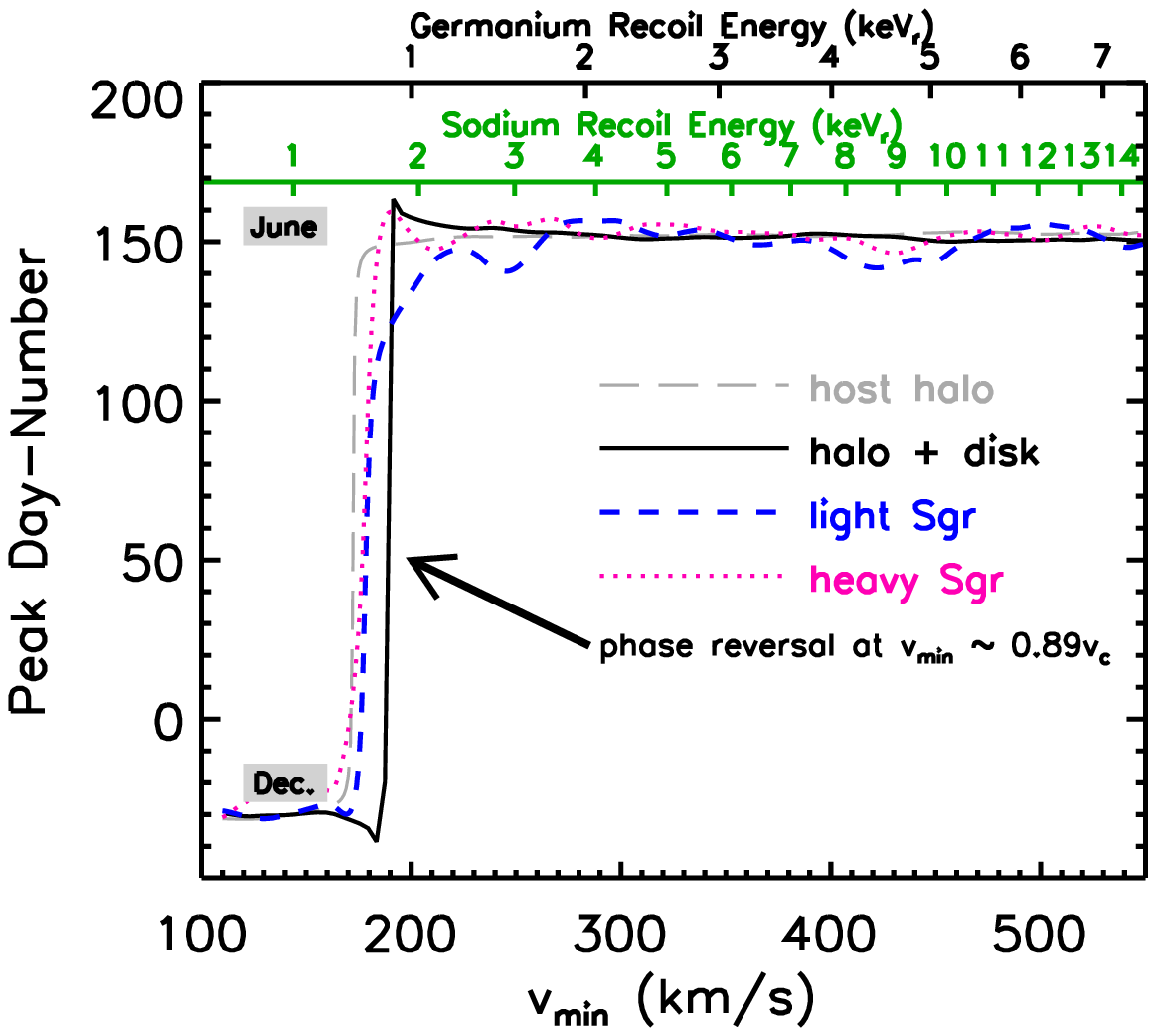}
\hspace{0.1in}
\includegraphics[width=2.8in]{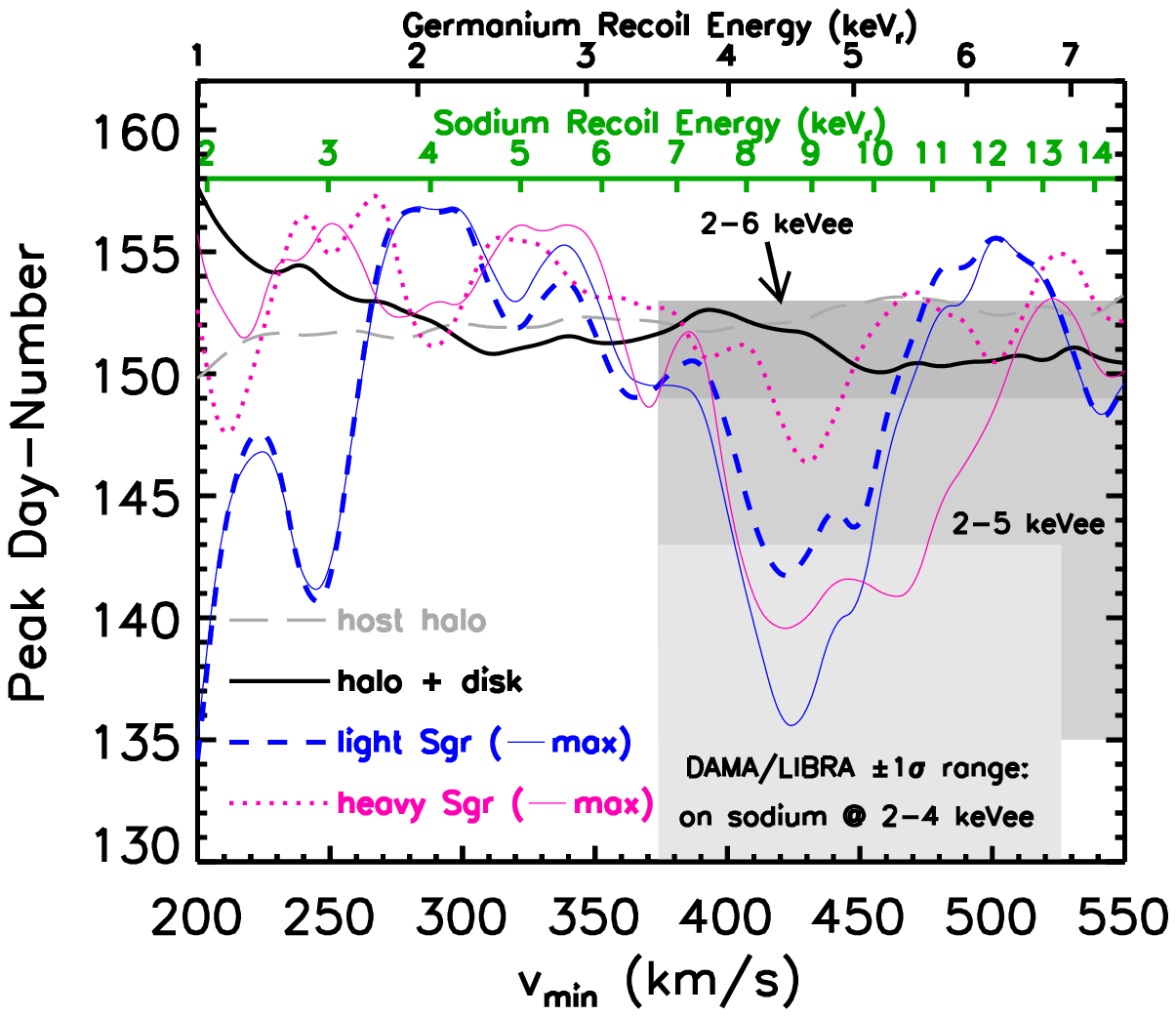}
\end{center}
\caption{Annual modulation of the dark matter scattering event rate.  
{\em Upper left} panel: at a fixed value of $v_{\mathrm{min}} \simeq 430$~km/s, corresponding to a recoil energy of 
$E_{\mathrm{r}} \simeq 4.6$~keV$_{\mathrm{r}}$ for a germanium detector, assuming a light-WIMP mass of $m_{\chi} = 10$~GeV$/c^2$, we show the best-fit amplitude $A$ and peak day-number for each model.  
{\em Upper right} panel: The fractional annual modulation amplitude as a function of $v_{\mathrm{min}}$, i.e. $\mathrm{(max-min)/(max+min)}$. 
{\em Lower} panels: The dependence of the 
peak day-number on the minimum velocity $v_{\mathrm{min}}$, with equivalent 
recoil energies for germanium- and sodium-based detectors plotted on alternate 
{\em upper} axes for $m_{\chi} = 10$~GeV$/c^2$.  
Note that the peak day-number shifts phase to Northern winter, 
for small unquenched recoil energies $E_{\mathrm{r}} \lesssim 1~(2)$~keV$_{\mathrm{r}}$ 
for germanium (sodium), corresponding to $v_{\mathrm{min}} \simeq 0.89v_c$ 
in agreement with ref.~\cite{lewis_freese04}.  In the {\em lower right} panel, 
the shaded regions represent the $\pm 1\sigma$ ranges in the modulation peaks determined 
by DAMA/LIBRA for scattering on sodium, i.e. peak day-numbers of $136,142,146 \pm 7$~for recoil energy bins $E_{\mathrm{r}} = 2-4,5,6$~keVee \cite{bernabei_etal11}.  Where present, thin colored lines 
represent the maximum possible signal induced by each Sgr model, as discussed in the text.
}
\label{fig:annmod}
\end{figure}

\subsection{Dark Matter from Sagittarius at Earth}
\label{subsec:sgr}

Our most novel results pertain to the effects of the Sagittarius debris on direct search rates.  Before 
exploring the influence of the Sagittarius tidal debris in detail, several points are worthy of reiterating 
here.  Our Sagittarius models are cosmologically-motivated and are designed to bracket the range 
of halo masses that a Sgr-like galaxy would be expected to have as it merged with the Milky Way.  
Both Sagittarius models produce debris streams that are in broad agreement with the known morphological 
and kinematical properties of the observed Sagittarius stream stars.  Moreover, as noted already in Fig.~\ref{fig:sgr}, 
in these self-consistent models of Sagittarius evolution, the dark matter stream is significantly 
wider spatially than the stellar stream, spanning $\gtrsim 10$~kpc in the direction transverse to the 
stream.  Lastly, the dark matter and stellar debris streams 
are not necessarily coincident; in fact, at the point where the Sagittarius stream penetrates 
the disk closest to the solar neighborhood, the dark matter and stellar streams are not 
coaxial.  The offset between the dark matter and stellar streams stems from the fact that the 
progenitor Sagittarius dark matter halo is significantly more extended than the stellar 
component of the progenitor dwarf galaxy, so that dark matter is typically liberated 
prior to stellar material.  Consequently, the dark matter and stellar streams do not follow the same 
orbits.  These considerations suggest that the dark matter component of the 
Sagittarius stream may likely be a very important contributor to scattering in earthbound 
direct detection experiments even if the stellar stream penetrates the Milky Way disk more than two kpc 
away from the Sun as is currently suspected (for example, in the modeling of ref.~\cite{law_etal10}, 
which incorporates the general observational results of \cite{seabroke_etal08,correnti_etal10}, among 
others).  These broad points represent an important addition to the literature on the influence of Sagittarius for 
dark matter direct detection experiments.  

The overall fraction of dark matter in the solar neighborhood is 
relatively small in all of our models, varying from $\sim 1-2\%$~of $\rho_0$ and may reasonably be as 
high as $\sim 5\%$~of $\rho_0$ if $\rho_0=0.3~\mathrm{GeV}/\mathrm{cm}^{3}$ rather than the somewhat higher 
values in our models.  Consequently, integrated event rates are altered by 
only relatively small amounts by the stream.  Figure~\ref{fig:veldistg} and 
Figure~\ref{fig:drde} show that the stream debris can alter event rates 
by up to $\sim 10-20\%$ at relative speeds greater than the typical 
relative stream speed, $v_{\mathrm{min}} \gtrsim 400$~km/s (if the 
local density from the host halo is as low as $\rho_0=0.3~\mathrm{GeV}/\mathrm{cm}^{3}$ 
as in the SHM, this boost can be as large as $\sim 25\%$).
Note that the Sagittarius debris give a relative enhancement to the speed 
distribution at speeds near $v \sim 400-500$~km/s in Fig.~\ref{fig:veldistg}.  
The best-fit Maxwellian is a better description of the models that include 
Sagittarius debris than it is of the {\em halo}$~+~${\em disk} models because 
the high-speed particles in {\em light}~and {\em heavy Sgr} influence the fitted 
kurtosis sufficiently that some of the halo's natural non-Maxwellianity is accounted for more closely.  

Rather than representing large modifications to the overall annually-averaged event 
rate, the speed distribution features around $400-500$~km/s due to the Sagittarius debris 
represent potentially interesting peculiarities that may be explored with the 
annual modulation of the event rate by current-generation experiments as well as directional detection 
efforts addressing the North Galactic Pole (latitude $b \sim +90^{\mathrm{o}}$ with 
respect to the solar neighborhood) in the long-term future.  
The stripped dark matter from Sgr falls coherently toward the solar system from the 
North Galactic Pole at a speed of order $\sim 400-500$~km/s.  The speed relative to 
the Earth peaks during Northern winter, nearly opposite in 
phase to the relative speed between the Earth and the dark matter in the primary host 
halo of the Milky Way, due to the geometry of the Sgr orbit with respect to the Milky Way disk plane.

We present results regarding the annual modulation of dark matter direct search rates 
in Figure~\ref{fig:annmod}.  In the {\em upper left} panel of Fig.~\ref{fig:annmod}, we 
show an example of the direct search event rate as a function of day-number (relative to noon 
on December 31st during the J2000.0 epoch) for a 
germanium detector, a light WIMP with mass $m_{\chi} = 10$~GeV$/c^2$, and 
unquenched nuclear recoil energy of $\simeq 4.6$ keV$_{\mathrm{r}}$.   
The amplitude of the annual modulation of 
$g(v_{\mathrm{min}})$ is shown in the {\em upper right} 
panel of Fig.~\ref{fig:annmod}.  We show the amplitude as the 
difference between the maximum and minimum event rates achieved within 
a yearly cycle divided by the sum of the maximum and minimum rates.  If the 
modulation induces a simple sinusoid superimposed upon a constant background, 
this yields the amplitude of the sinusoid.  The amplitude of the annual modulation 
signal is generically an increasing function of $v_{\mathrm{min}}$ because the number of 
particles in the tail of the speed distribution is very sensitive to small shifts in the 
central position of the speed distribution.  The SHM predicts a much more shallow rise in the 
fractional modulation amplitude at relatively high speeds than that yielded by all four of our models, 
as shown in the {\em upper right} panel of Fig.~\ref{fig:annmod}.  
Each of the models containing a stellar disk results in a fractional amplitude that is 
more than $20\%$ lower than that of the {\em host halo} model alone.  

The Sagittarius debris alters predictions for the amplitude of the annual modulation markedly.  In particular, both of our 
fiducial models of Sagittarius ({\em dashed lines}) exhibit an annual modulation amplitude 
as much as $\sim 20-30\%$ less than the modulation amplitude in the Galaxy models with no 
Sagittarius debris (and a factor of $\sim 2$ smaller than in the {\em host halo}) in the range 
$400~\mathrm{km/s} \lesssim v_{\mathrm{min}} \lesssim 500~\mathrm{km/s}$.  
The modulation amplitude is reduced in the Sagittarius models because the dark matter stream 
of Sagittarius rains down upon the Galactic disk from the North Galactic direction, so that 
this component is distinctly out of phase with the background ``wind'' of dark matter 
particles from the primary Galactic halo \cite{freese_etal04,freese_etal05,savage_etal06,savage_etal07}.

As we have alluded to previously, the location of the impact of the leading stellar stream of the Sagittarius 
dwarf on the Milky Way disk relative to the solar neighborhood remains uncertain, 
partly because high-latitude data must be extrapolated to the disk plane in 
order to estimate the impact position \cite{correnti_etal10}, and partly due to uncertainty in the Sun's position.  Consequently, 
we have estimated the possible range of annual modulation effects that may 
be attributable to the Sagittarius debris by artificially shifting the 
position of the Sagittarius impact with respect to the Sun in our models 
by amounts consistent with observational constraints on the relative position 
of the Sagittarius impact with the disk and the Sun.  
As shown in the {\em lower panels} of Figure~\ref{fig:sgr}, 
the solar neighborhood in the {\em light Sgr} model is serendipitously
near the peak of the Sagittarius dark matter density 
(despite the {\em stellar} stream being several kpc away, because 
the central axes of the dark and luminous tidal arms are not coincident), 
while the {\em heavy Sgr} model's solar position is $\sim 2-3$~kpc farther from 
the Galactic Center than the axis of the dark stream.  Our models suggest 
that if the Sun is $\sim 10$~kpc or more from the center of the {\em stellar} stream 
impact on the Galactic plane, the influence of Sagittarius on direct search experiments may be quite 
small; however, current modeling and observational constraints indicate that the leading arm is 
probably significantly closer to the solar neighborhood \cite{correnti_etal10}.  

To probe this uncertainty, we estimate a range of possible 
Sgr-related effects by taking the maximal local dark matter density 
commensurate with observational determinations of the {\em stellar} stream 
position.  To a good approximation, the maximal influence of Sagittarius 
occurs when the leading {\em dark matter} arm falls directly onto the solar 
position.  This can be achieved in both models even when the peak density 
of the {\em stellar} arm is more than a kiloparsec from the Sun (see Fig.~\ref{fig:sgr}).  
Moreover, we note that although observations indicate that the 
solar neighborhood is not significantly contaminated by former Sagittarius 
stars \cite{helmi_etal99,re_fiorentin_etal11}, kinematically-streaming
sub-populations that may be associated with Sagittarius would only 
contribute on the order of $\sim 1\%$ to the local 
stellar density, making them very difficult to rule out on heliocentric-distance scales larger than a 
kiloparsec.   We refer to the cases in which 
the relative position of the Sun maximizes the annual modulation influence 
of Sagittarius as our {\em maximal} models for the {\em heavy Sgr} and {\em light Sgr} 
simulations respectively.  

The maximal Sagittarius cases are shown by {\em thin, solid} lines in the 
{\em upper right} panel of Fig.~\ref{fig:annmod}.  The additional 
decrease in the annual modulation amplitude is significant.  In particular, 
the annual modulation amplitude can be decreased by as much as a factor 
of two in the maximal Sagittarius models compared to the {\em halo}$~+~${\em disk} 
models in the speed range $400~\mathrm{km/s} \lesssim v_{\mathrm{min}} \lesssim 500~\mathrm{km/s}$.

The Sagittarius stream dark matter has an important influence on the phase of the annual 
modulation of scattering rates.  The phase of the annual modulation as a function of 
$v_{\mathrm{min}}$, specified by the day of the peak signal (with day-number $1$ set to noon on December 31st in the J2000.0 epoch), 
is shown in both {\em lower} panels in Fig.~\ref{fig:annmod}.  
The peak days in the {\em host halo} and {\em halo}$~+~${\em disk} models are 
close to the canonical value of day 152.5 for $v_{\mathrm{min}} \gtrsim 190$~km/s.  
At lower values of $v_{\mathrm{min}}$, the phase switches to a peak day in 
the Northern winter, close to day 335 (or equivalently day $-$30 as shown in Fig.~\ref{fig:annmod}).  
The shift in the phase of the oscillation at low energies occurs because the shift to 
higher relative speeds caused by the greater relative speed of the Earth compared to the 
primary host halo dark matter particles during Northern summer augments the high-speed 
portion of the speed distribution, but depletes the low-speed portion of the speed distribution 
so that low-energy scattering rates are reduced during Northern summer.  Conversely, low-energy 
scattering rates are enhanced during Northern winter.  For the SHM, the peak day at high 
$v_{\mathrm{min}}$ occurs on day 152.5 (June 2nd), while the peak day at low $v_{\mathrm{min}}$ 
occurs on day 335 (December 1st).  The shift of the peak day from Northern winter to Northern summer 
occurs at $v_{\mathrm{min}} \simeq 0.89 v_c$ in the SHM \cite{lewis_freese04}.  All of our models show a similar 
phase reversal in the peak day, though the value of $v_{\mathrm{min}}$ at which the shift occurs varies slightly 
with respect to the SHM value, as we discuss shortly in the context of the Sagittarius models.

\begin{figure}
\begin{center}
\includegraphics[width=4in]{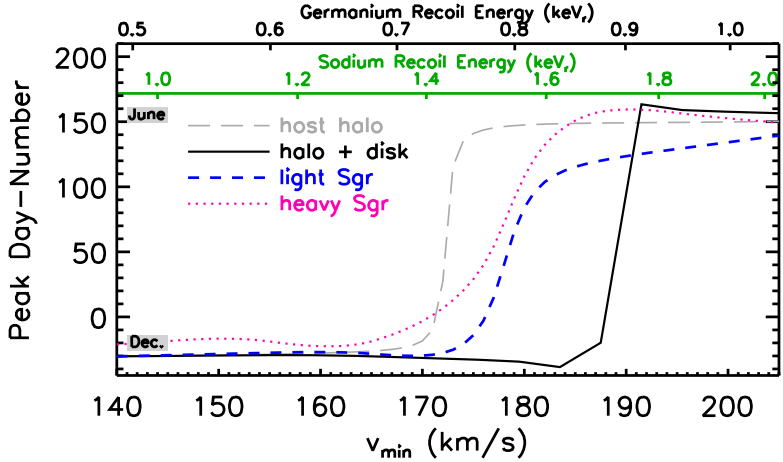}
\end{center}
\caption{The phase reversal of the annual modulation signal at relatively low recoil energies, 
for each of our four models.  Note that the peak occurs in Northern winter for $v_{\mathrm{min}} \lesssim 0.89v_c$ 
as expected\cite{lewis_freese04}, for the {\em host halo} and {\em halo}$~+~${\em disk} models.  
The Sagittarius models reverse phase at lower $v_{\mathrm{min}}$.  
This discrepancy could potentially manifest itself following a 
large number of scattering events observed by the next generation of 
direct detection experiments, which may feature sub-keV recoil 
energy thresholds and improved energy resolution.  As in other figures, equivalent recoil energies for 
germanium- and sodium-based detectors are plotted on alternate {\em upper} axes for $m_{\chi} = 10$~GeV$/c^2$.  
Heavier WIMPs, near the canonical mass range of $m_{\chi} \sim 10^2 - 10^3$~GeV, imply higher recoil energies 
than those shown on the upper axes here.
}
\label{fig:lowenergy}
\end{figure}

The Sagittarius streams in our simulations alter the annual modulation signals 
in non-negligible ways.  The differences between the {\em halo}$~+~${\em disk} or {\em host halo} 
models as compared to the Sagittarius models depend upon $v_{\mathrm{min}}$.  The most 
significant deviations are for $v_{\mathrm{min}} \lesssim 250$~km/s and the range of 
speeds typical of the relative speed of the Earth with respect to the Sagittarius stream, 
$400~\mathrm{km/s} \lesssim v_{\mathrm{min}} \lesssim 500~\mathrm{km/s}$.  In the higher 
speed range, the maximal shifts in peak day bring the peak between 5 and 12 days earlier 
in the year for the fiducial {\em heavy Sgr} and {\em light Sgr} models respectively.  
In the maximal models, the shift can be larger, bringing the peak day as much as 
12 and 18 days earlier in narrow energy ranges in the {\em heavy Sgr} and 
{\em light Sgr} models respectively.  The shifts in peak day are comparable in 
magnitude to the errors on the DAMA/LIBRA measurement of annual modulation \cite{bernabei_etal11}.  
Moreover, the shifts in our models bring the peak day 
earlier by amounts and with energy dependence comparable to the offset between the 
SHM peak day and the peak day quoted by the DAMA/LIBRA experiment, for which the signal 
peaks at day-number $136,142,146 \pm 7$ in the recoil-energy bins $2-4,5,6$~keVee, 
as indicated by the shaded regions in the lower right panel of 
Figure~\ref{fig:annmod}.  Additionally, the energy dependence of the annual modulation amplitude 
in our models may partially mitigate the discrepancy between the low amplitude observed by DAMA/LIBRA at energies $2-6$~keVee 
($A \sim 2\%$ of the mean rate 0.0116 cpd/kg/keVee) and the somewhat higher amplitude $A \sim 13\%$ quoted by CoGeNT at energies $0.5-3.0$~keVee.  
Conversely, the peak day shift induced by 
the Sgr debris becomes indistinguishable from noise, if the solar position 
can be located as much as $\sim 5$~kpc from the center of the {\em stellar} 
debris arm, a distance somewhat greater than contemporary estimates. 

It is tempting to suggest that experiments such as 
DAMA/LIBRA or CoGeNT may be probing the stream of the Sagittarius galaxy, but any such 
statement is subject to numerous important caveats.  First, the DAMA/LIBRA 
results are controversial and are challenged by other experiments.  Second, 
the DAMA/LIBRA peak day estimates deviate from the SHM (and {\em halo}$~+~${\em disk} model) 
predictions by only $2\sigma$ or less.  However, another caveat to any 
such statement strikes at a limitation of our simulation analysis.  
As we have already mentioned, the position of the Sagittarius stream 
impact on the disk remains uncertain.  Further the angle at which the 
Sagittarius stream impacts the disk is also uncertain.  Finally, despite the fact 
that these simulations are the most cosmologically complete of such efforts, having accounted for 
realistic dark matter halos in each galaxy, some small discrepancies remain between our 
results and observations of the sky-position and radial velocity of the stellar debris.  Unfortunately, 
a parameter search using numerical simulations is not yet practicable.  As an example of the concrete 
effects of this uncertainty on our predictions, the amount 
of the phase shift induced by the Sagittarius stream and, in fact, 
whether the shift is toward earlier in Northern spring (as shown in 
Fig.~\ref{fig:annmod}) or, perhaps, later in Northern summer is sensitive 
to the angle at which Sagittarius impacts the disk.  
Contemporary data and techniques do not suffice to specify this 
impact direction, so it is not possible to interpret our results 
as a firm statement that the Sagittarius stream causes the peak 
day to occur earlier by some precise amount (the amplitude of the annual modulation is 
not subject to this caveat).  A proper interpretation of our simulation 
results is that we have shown that it is possible, in 
self-consistent models designed to mimic the Milky Way with 
the Sagittarius impact, for the Sagittarius debris to induce 
significant, energy-dependent shifts in the annual modulation phase 
even when the {\em stellar} stream of the Sagittarius debris 
remains several kiloparsecs from the Sun.   

We now turn to the phase reversal at low velocities/recoil energies.  As we have 
already mentioned, the phase reversal 
of the annual modulation at low energies is a well-understood effect, 
occurring at $v_{\mathrm{min}} \simeq 0.89v_c$ in the SHM \cite{lewis_freese04} and at similar 
values in our {\em host halo} and {\em halo}$~+~${\em disk} models.  
As depicted in Figure~\ref{fig:lowenergy}, the presence of Sgr dark matter 
in the solar neighborhood affects not only the phase of the overall modulation, but also the 
value of $v_{\mathrm{min}}$, and thus the recoil energy, 
at which the phase reversal occurs.  In particular, the Sagittarius 
models reverse phase at speeds $\sim 10-15$~km/s lower than the 
{\em host halo} and {\em halo}$~+~${\em disk} models.  For relatively 
low-mass WIMPs with $m_{\chi} = 10$~GeV, this corresponds to a shift in recoil energy 
at the phase reversal point of approximately $\Delta E_{\mathrm{r}} \sim 0.1-0.2$~keV$_{\mathrm{r}}$.  
For larger WIMP masses in the regime $m_{\chi} \gtrsim 100~\mathrm{GeV/}$, this shift 
in the reversal energy would be larger by a factor of $10$ or more.

This Sagittarius-induced feature may be probed by future detectors 
with low energy threshold and improved energy resolutions and may be 
one of the distinguishing features of Sagittarius if WIMP astronomy 
can ever be undertaken in an era with very large direct search 
event rates (as explored by ref.~\cite{peter2011}).  
Exploring this particular signature is well suited for efforts to 
develop very low-threshold direct search detectors with greatly 
improved energy resolution, examples of which include extensions of 
germanium-based experiments like MAJORANA \cite{barbeau_phd09,phillips_etal11}, 
advanced threshold-lowering and background-eliminating technologies like those proposed 
by the CDEX-TEXONO and SuperCDMS/CDMSLite collaborations (as in refs.~\cite{texono,bruch2010,pyle_etal12}, respectively).  
This avenue of exploration may be particularly fruitful in the 
future as the local circular speed $v_c$ of the Milky Way is refined by next-generation 
astrometric surveys so that a similarly tight constraint can be placed on 
the phase-reversal energy predicted by the SHM and similar models without 
Sagittarius debris.

There is a final addendum to our model results that is worth stating explicitly.  In our models, the 
potential interior to the solar position is dominated by stellar mass, rather than dark matter.  
In the event that the local dark matter density contributed by the primary host halo of the Milky Way 
(as opposed to the Sagittarius stream) is as low as the SHM value of $\rho_0 \sim 0.3~\mathrm{GeV/cm^3}$, 
we expect viable models of Sagittarius evolution to remain broadly similar.  The implication is that 
the {\em relative} influence of Sagittarius on direct search scattering rates could be yet larger than 
we have estimated here.  Although we have not performed a self-consistent model that results in 
such parameters for the Milky Way halo, it is interesting to comment on how our results would 
change in such a scenario.  If the local density contributed by the host halo of the Milky Way 
were as low as $\rho_0 \sim 0.3~\mathrm{GeV/cm^3}$ (rather than roughly twice this value as in 
our equilibrium halo/galaxy models), the Sagittarius debris could contribute as much as $\sim 5\%$ 
of the local dark matter density and the direct search signatures we have explored would change 
as follows: the phase-shift in the annual modulation signal could be as large as $\sim 20-25$ days, compared 
to the $\sim 10$-day shift yielded by our fiducial models, and the fractional amplitude of the annual modulation signal 
could remain as low as $\sim 5\%$ even at relatively high speeds near $v_{\mathrm{min}} \sim 500$~km/s 
(well below the SHM prediction as well as each of the cases we examine here).  In addition, the change in the 
recoil energy at which the modulation undergoes phase reversal could be as much as twice the fiducial 
change we show in Fig.~\ref{fig:lowenergy}.  These estimates, based on a scenario in which the solar neighborhood 
is near the peak of the dark matter stream {\em and} the local density contributed by the parent halo is as low as 
$\rho_0 = 0.3~\mathrm{GeV/cm^3}$, likely represent maximum plausible influences that Sagittarius stream material 
could have on direct search rates without fine-tuning.

 \section{Discussion}
\label{sec:discuss}

We have studied predictions for dark matter direct search scattering rates within the context 
of isolated numerical models of a Milky Way-like system designed to reproduce the basic properties 
of the Galaxy, including models of the infall, merger, and tidal disruption of the Sagittarius 
dwarf system.  In modeling such specific features, isolated simulations 
of this kind complement large-scale cosmological simulations.  
In accord with previous high-resolution studies of 
cosmological dark matter halos 
\cite{diemand_etal04,vergados_etal08,fairbairn_etal09,kuhlen_etal10}, 
we find that deviations from standard halo model (SHM) assumptions in observationally 
viable model Milky Way systems can significantly alter direct search rates relative to SHM 
predictions.  In agreement with these studies, equilibrated host halo systems and halo systems 
in equilibrium with a Galaxy exhibit speed distributions that differ markedly from the 
Maxwellian form, in particular being significantly platykurtic.  

Not surprisingly, in all three models that include an equilibrated Galactic stellar disk, 
we find that the increased relative speeds of dark matter particles caused by the additional 
acceleration provided by the disk and the contracted halo result in significantly larger 
scattering rates.  The precise value of this enhancement can be large (a factor of several) 
and is energy dependent (Fig.~\ref{fig:veldistg} and Fig.~\ref{fig:drde}).  
Studies based upon $N$-body realizations of Galaxy-analog halos in cosmological simulations 
make the implicit assumption that a Galaxy-sized halo in a cosmological numerical experiment 
involving only dark matter will faithfully reflect the solar neighborhood in the real 
Galaxy \cite{vergados_etal08,fairbairn_etal09}.  For practical purposes, mapping cosmological 
N-body results onto equilibrium models containing a galaxy is non-trivial.  We have shown 
that our equilibrium {\em halo}$~+~${\em disk} models can be used to predict rates to within 
50\% by contracting the dark matter halo using standard adiabatic contraction techniques 
 \cite{blumenthal_etal86,gnedin_etal04,binney_tremaine08}.  
Meanwhile, simply scaling the speed distribution of the $N$-body host halo 
by the mean velocity can reproduce the {\em halo/galaxy} distribution to within $\sim 5\%$ precision, 
signifying that dark matter-only predictions for speed distributions can be mapped to models that include a 
Milky Way Galaxy by scaling speeds up to the rotation speed in the solar neighborhood.  
These results are broadly commensurable with previous $N$-body work 
and a significant caveat to these results is that our equilibrium models represent only one possible equilibrium 
solution for the halo/galaxy system that is not unique and does not result from self-consistent cosmological evolution.

Our most novel results pertain to the influence of Sagittarius debris material on 
predicted direct search event rates.  Our models demonstrate that the Sagittarius 
stream debris can have an important influence on direct search scattering rates 
even when the {\em stellar} stream of the Sagittarius debris is centered several 
kiloparsecs from the solar neighborhood, as is thought most likely based on 
contemporary analyses \cite{seabroke_etal08,law_etal10,correnti_etal10}.  
The reasons for this are twofold.  First, the Sagittarius dark matter stream is significantly 
broader than the stellar stream in our models.  The Sagittarius stream is a non-negligible 
contribution to the nearby dark matter content over an area many kiloparsecs in diameter 
in the plane of the Galactic disk (Fig.~\ref{fig:sgr}).  Second, the Sagittarius 
stellar and dark matter streams are not spatially coincident in our models, having drifted 
away from co-axiality during evolution in the Milky Way's tidal field.  The 
peak of the dark matter density contributed by the stream impacts the Galactic 
disk several kiloparsecs from the peak of the stellar density, and the spread in surface 
density demonstrated by Fig.~\ref{fig:sgr} indicates that expected event-rate boosts for 
detection experiments should be important out to this distance, in our adopted formalism.

We reiterate an important caveat before continuing; although the spatial and kinematic 
distributions of Sagittarius stellar debris in our models are generally good matches to the 
observational properties of that debris, as elaborated in \cite{purcell_etal11}, some discrepancies 
do remain.  Moreover, the relative position of the solar neighborhood with respect to the Sagittarius 
stream's impact point on the Galactic disk is still poorly constrained.  The simulations we analyze are 
among the most complete descriptions of the Sagittarius debris in the literature and an exhaustive search 
of the initial parameter space is not computationally feasible, so these uncertainties cannot be 
explored in detail.  A proper interpretation of our results would be that the Sagittarius dark matter 
debris may give rise to significant signals in direct-search experiments even if the Sagittarius stellar 
debris is confined to a distance of several kiloparsecs from the solar neighborhood. 
Tuning the stellar stream more finely would not impact our results, insofar as  
variation in Sgr WIMP surface-density at the Sun is less important to the event-rate calculation 
than is the vertical velocity distribution of those WIMPs.  The important and general implication of this result is 
that the stellar stream of Sagittarius being several kiloparsecs from the solar neighborhood 
{\em does not} preclude Sagittarius from having a significant effect on dark matter 
experiments.  Furthermore, near-future constraints on the location of the 
stellar debris may not suffice to preclude significant dark matter 
from Sagittarius in the solar neighborhood.

The Sagittarius stream gives rise to several important effects on dark matter 
experiments.  The high relative velocity of the Sagittarius dark matter 
stream relative to the Earth boosts the rate of high-energy recoil events 
by $\sim 20\%$ in our models and perhaps as much as $\sim 40-45\%$ depending 
upon the local dark matter density contributed by the primary halo of the 
Milky Way (Fig.~\ref{fig:veldistg} \& Fig.~\ref{fig:drde}).  
Sagittarius reduces the annual modulation amplitude by an energy-dependent 
factor that may be as large as a factor of two at $v_{\mathrm{min}} \sim 420$~km/s 
(Fig.~\ref{fig:annmod}) relative to models with no Sagittarius debris.  
This energy-dependent suppression could help explain the disparate values found 
by DAMA/LIBRA (where the modulation amplitude is $A \sim 2\%$ of the 
mean low-energy rate $dR/dE \simeq 0.0116$~cpd/kg/keVee in the energy bin $2-6$~keVee~$=6.7-20$~keV$_\mathrm{r}$) 
and CoGeNT (having an amplitude $A \sim 13\%$ of the mean rate for recoil energies $\sim 0.5-3.0$~keVee~$=2.3-11.3$~keV$_\mathrm{r}$) 
as well as their discrepancies compared to the amplitudes expected within the SHM 
\cite{bernabei_etal11,aalseth_etal11}, because our Sagittarius models result in 
amplitudes that increase much more sharply with recoil energy in this range than 
the SHM formalism as well as the {\em host halo} and {\em halo}~+~{\em disk} models.

The geometry of the Sagittarius impact on the solar system 
causes the signal from the Sagittarius stream to peak during 
Northern winter, in agreement with previous studies 
\cite{freese_etal04,freese_etal05,savage_etal06,savage_etal07,savage_etal09} 
(as well as the general study of debris flows in 
ref.~\cite{vergados_2012}).  Our models of Sagittarius 
yield recoil energy-dependent shifts in peak day number of 
between $\sim 5$~and $25$ days earlier in the year than the SHM peak day-number of $152.5$.  
Both DAMA/LIBRA and CoGeNT have indicated a similar behavior, 
with both experiments finding trends between peak day-number and 
recoil energy \cite{bernabei_etal11,aalseth_etal11}; however, 
the observational error in the peak remains on the order of a few days 
and the uncertainty in simulation programs that aim to model Sagittarius 
remain significant.  Nevertheless, our models suggest such a shift 
is reasonable given contemporary knowledge of Sagittarius debris structure.
The Sagittarius streams in our models also cause the phase reversal 
energy to be lowered (Fig.~\ref{fig:lowenergy}).  
Exploiting this signature, in particular, to help identify 
dark matter or use dark matter searches to perform WIMP 
astronomy will benefit greatly from future 
low-threshold detectors with improved energy resolution, 
such as those being considered by refs.~\cite{texono,bruch2010,pyle_etal12}.  

There have been a number of previous studies of the influence of Sagittarius 
debris on direct dark matter searches, including 
refs.~\cite{freese_etal04,freese_etal05,savage_etal06,savage_etal07,savage_etal09,natarajan_etal11}.  
By and large, these studies have utilized the SHM formalism and 
contrived approximations for the local distribution of Sagittarius 
debris based on observational limits on the Sagittarius {\em stellar} debris 
contribution to the solar neighborhood.  Further, the prospects for Sagittarius to significantly 
influence direct search experiments have been challenged as contemporary data 
suggest that the {\em stellar} component of the Sagittarius debris is centered 
a few kpc from the solar neighborhood \cite{law_etal10,correnti_etal10} (and here 
we reiterate that although refs.~\cite{helmi_etal99,re_fiorentin_etal11} find no evidence of {\em large} coherent 
sub-populations in the solar vicinity, percent-level debris flows are presently unconstrained beyond one or two 
kiloparsecs from the Sun).
Our study complements previous work in many respects.  Most importantly, 
we have analyzed a self-consistent simulation of Sagittarius dwarf galaxy accretion 
involving a realistic dark matter component, doing so in a controlled and isolated simulation 
such that event-rate implications can be clearly discerned.  
Crucially, we have shown that the proximity of the Sun 
to the {\em stellar} stream alone cannot necessarily be used 
as an indicator of the local dark matter contribution from 
Sagittarius, as the dark matter flow accompanying the luminous debris 
is much more widely spread across the Milky Way disk and not necessarily 
coincident with the stellar material.

Comparing to previous work in more detail, we present the first analysis of the influence of Sagittarius dark matter 
based on self-consistent models of the Sagittarius infall that describe the observed Sagittarius debris within observational 
uncertainty.  Unlike previous studies, we emphasize that the Sagittarius debris induces an increase in the event rate by $\sim 10-20\%$ in 
our fiducial models (as much as $\sim 40\%$ in our maximal models), and that annual modulation fractional amplitudes are diminished by 
$\sim 20\%-50\%$ in the presence of that debris at Earth.  The phase shifts in the annual modulation that we 
find are somewhat larger than those presented by refs.~\cite{freese_etal04,freese_etal05} (for comparable dark matter density contributions), 
due to the relatively higher speed attained by Sgr particles in our modeling.  We find qualitatively similar phase behaviors to those of 
refs.~\cite{savage_etal06,savage_etal07}, near the reversal point at which the modulation amplitude changes sign.  Generally, our work 
agrees well with past commentary on the detectability of WIMP streams in current- and next-generation detection efforts 
\cite{natarajan_etal11,kuhlen_etal12,savage_etal09}, and specifically we identify the Sagittarius dark matter stream as 
an achievable target for direct-search science over the next decade.

The effects of Sagittarius that we describe in this manuscript may be relevant to 
dark matter searches generally.  However, if the dark matter is 
indeed relatively light ($m_{\chi} \lesssim 20$~GeV as we have assumed in 
our illustrative examples), the effects of Sagittarius debris on 
scattering rates are particularly important because of the large 
relative speed of the debris stream at the Earth.  In either case, 
future direct search experiments may probe such signatures, though a 
future generation of low-threshold detectors with fine energy 
resolution \cite{texono,bruch2010,pyle_etal12,graham_etal12}
may be necessary in the event that the dark matter mass falls in this 
lower range.  In either case, our analysis suggests that the effects of Sagittarius 
debris on direct search experiments {\em will not} be negligible given 
contemporary limits on the position of the Sagittarius {\em stellar} stream.  
In the far future, the features induced by Sagittarius debris may be among the early 
measurements to be made in an era of WIMP astronomy with large direct search rates \cite{peter2011}.

\acknowledgments{We would like to thank Larry Widrow, John Dubinski, 
and Miguel Rocha for kindly making available the software used to 
set up the initial galaxy models.  We also thank 
Andrew Hearin and Arvind Natarajan for useful discussions.  This work was 
initiated during the "Exploring Low-Mass Dark Matter Candidates" workshop organized 
by PITT PACC in November 2011 and we are grateful to the workshop participants for 
encouragement and numerous stimulating discussions.    
This work was funded by the Pittsburgh Particle physics, Astrophysics, 
and Cosmology Center (PITT PACC) and by the National Science Foundation 
through Grant NSF-PHY-0968888.}

\bibliographystyle{jhep}
\bibliography{dmdf}

\end{document}